\documentclass[a4paper,fleqn,usenatbib]{mnras}

\usepackage[T1]{fontenc}
\usepackage{ae,aecompl}
\usepackage{textcomp}
\usepackage{graphicx}	% Including figure files
\usepackage{amsmath}	% Advanced maths commands
\usepackage{amssymb}	% Extra maths symbols
\usepackage{multicol}        % Multi-column entries in tables
\usepackage{bm}		% Bold maths symbols, including upright Greek
\usepackage{pdflscape}	% Landscape pages
\usepackage{color}          % To give different colours for text
\newcommand{\mm}{\mu m}

\title[MIR $PL$ relations of RRLs in Reticulum]{The Carnegie RR Lyrae Program: Mid-infrared Period-Luminosity relations of RR Lyrae stars in Reticulum}

\author[T. Muraveva et al.]{Tatiana Muraveva$^{1}$\thanks{tatiana.muraveva@oabo.inaf.it}, Alessia Garofalo$^{1,2}$, Victoria Scowcroft$^{3,4}$, Gisella Clementini$^{1}$, 
\newauthor{Wendy L. Freedman$^{5}$, Barry F. Madore$^{4,5}$, Andrew J. Monson$^{6}$} 
\\
$^{1}$ INAF-Osservatorio di Astrofisica e Scienza dello Spazio di Bologna, Via Piero Gobetti, 93/3, Bologna 40129, Italy\\
$^{2}$ Universit\'a  di Bologna, Dipartimento di Fisica e Astronomia, Via Piero Gobetti 93/2, Bologna 40129,  Italy\\
$^{3}$ Department of Physics, University of Bath, Claverton Down, Bath, BA2 7AY, UK \\
$^{4}$ Observatories of the Carnegie Institution of Washington, 813 Santa Barbara St., Pasadena, California, CA 91101, USA\\
$^{5}$ Department of Astronomy and Astrophysics, University of Chicago, 5640 S Ellis Ave, Chicago, IL 60637, USA \\
$^{6}$ Department of Astronomy and Astrophysics, The Pennsylvania State University, 403 Davey Lab, University Park, PA, 16802, USA \\
}

\date{Accepted . Received ; in original form }

\pubyear{2018}

\begin{document}
\label{firstpage}
\pagerange{\pageref{firstpage}--\pageref{lastpage}}
\maketitle

\begin{abstract}
We analysed 30 RR Lyrae stars (RRLs) located in the Large Magellanic Cloud (LMC) globular cluster Reticulum that were observed in the 3.6 and 4.5 $\mm$ passbands with the Infrared Array Camera (IRAC)  on board of the {\it Spitzer Space Telescope}. We derived new mid-infrared (MIR) period-luminosity ($PL$) relations. The zero points of the $PL$ relations were estimated using the trigonometric parallaxes of five bright Milky Way (MW) RRLs measured with the {\it Hubble Space Telescope (HST)} and, as an alternative, we used the trigonometric parallaxes published in the first {\it Gaia}  data release (DR1) which were obtained as part of the Tycho-{\it Gaia} Astrometric Solution (TGAS) and the parallaxes of the same stars released with the second {\it Gaia} data release (DR2). We determined  the distance to Reticulum using our new MIR $PL$ relations and found that distances calibrated on the TGAS and DR2 parallaxes are in a good agreement and, generally, smaller than distances based on the {\it HST} parallaxes, although they are still consistent within the respective errors. 
%, however, the sample of  RRLs used in the analysis is too small to make a conclusion about the possible systematic effect. 
 We conclude that  Reticulum is located $\sim 3$~ kpc closer to us than the barycentre of the LMC.  
\end{abstract}


%%%%%%%%%%%%%%%%%%%%%%%%%%%%%%%%%%%%%%%%%%%%%%%%%%

%%%%%%%%%%%%%%%%% BODY OF PAPER %%%%%%%%%%%%%%%%%%

\section{Introduction}

Knowledge of the distances to celestial objects is crucially important to all branches of astronomy.  In order to estimate  distances, astronomers have developed a number of different techniques. Some methods, such as the trigonometric parallax, are based on geometric  principles, while others invoke the use of  primary distance indicators such as Cepheids or RR Lyrae (RRL) variables,  which in turn are calibrated using the geometric methods. RRL stars (RRLs)  are radially pulsating variables which are abundant in globular clusters (GCs) and in the halos of galaxies. The RRLs are old  ($>10$ Gyr),  low-mass ($\sim0.6 - 0.8$ $M_\odot$) core-helium burning stars that lie within the instability strip crossing the horizontal branch (HB) of the colour-magnitude diagram (CMD). They are divided into those pulsating in the fundamental (RRab) and first-overtone (RRc) modes, and both modes simultaneously (RRd).
 RRLs are an excellent tool to  estimate distances in the Milky Way (MW) and to Local Group galaxies because they conform to an optical luminosity-metallicity ($M_{V}-{\rm [Fe/H]}$) relation and  to infrared period-luminosity ($PL$)  and $PL$-metallicity ($PLZ$) relations.
The  near-infrared (NIR) and mid-infrared (MIR) $PL$ relations of RRLs have several advantages in comparison with the visual $M_{V}-{\rm [Fe/H]}$ relation; specifically, a  milder dependence of the luminosity on metallicity and lower sensitivity to interstellar extinction.  Indeed, extinction in the 3.6 and 4.5 $\mm$ bands is much reduced with respect to the visual passband ($A_{3.6} \sim 0.06A_V$ and $A_{4.5} \sim 0.05A_V$, \citealt{Monson2012}). 
Furthermore, the  intrinsic scatter of the RRL (and Cepheid) $PL$ relations  decreases with increasing the wavelength \citep{Madore2012} and  %for RRLs and Cepheids .
the MIR light curves of RRLs (and Cepheids) have smaller amplitudes than in the optical passbands, hence, the determination of the mean magnitudes is easier and more precise. 
 Moreover, a number of studies also indicate that the  $M_V - {\rm [Fe/H]}$ relation may be not linear  (\citealt{Caputo2000}; \citealt{Rey2000}; \citealt{Bono2003}), which is not the case for the MIR $PL$ relations.

In order to study the RRL $PL$ relation in the MIR passbands it is necessary to have a statistically significant sample of variables with firmly established classification,  well-known periods and accurately determined mean MIR magnitudes. 
%In this study we used data obtained as part of the Carnegie RR Lyrae Program (CRRP), 
%% designed to calibrate the MIR $PL$ relations of RRLs in order to study the three-dimensional structure of the local universe and eventually to contribute to the determination of the Hubble constant (\citealt{Freedman2012}, PID 90002). 
%which collected observations  in the 3.6 and 4.5 $\mu m$ passbands with the Infrared Array Camera (IRAC, \citealt{Fazio2004}) on the {\it Spitzer Space Telescope} \citep{Werner2004} 
%as part of the Warm Spitzer Cycle 9  (PI W. Freedman), for samples of RRLs in the MW field, in a number of globular clusters and in a few dwarf spheroidal MW satellites.
%%executed during the warm mission of the {\it Spitzer Space Telescope} \citep{Werner2004} Infrared Array Camera (IRAC, \citealt{Fazio2004}) . 
In particular, for the present  study of the MIR RRL $PL$ relation we selected Reticulum, an old GC located in the Large Magellanic Cloud (LMC), $\sim$ 11 degrees away from the centre of its host galaxy \citep{Demers1976}. 
The RRL population of Reticulum has been studied by several  authors. \citet{Demers1976} discovered 22 RRLs in the cluster. \citet{Walker1992} later identified 10 additional RRLs  bringing  the  total number of RRLs in Reticulum to 32   (22 RRab, 9 RRc and 1 candidate RRd star). \citet{Ripepi2004} combined \citet{Walker1992}'s photometry with their observations  in the {\it B} and {\it V} passbands and detected double-mode behaviour in four of the previously discovered RRL variables. 

NIR $K$-band light curves with approximately 12 phase points per source  were published  by \citet{DallOra2004} for 30 of the RRLs in Reticulum.
Mean $K$-band magnitudes were estimated by fitting the light curves with templates from \citet{Jones1996} and then used to build a NIR $K$-band $PL$ relation.

Most recently, in their optical study of Reticulum, \citet{Kuehn2013} identified a secondary pulsation mode in two RRLs changing their classifications from RRc to RRd. Hence, the RRLs  in Reticulum now list 22 RRab, 4 RRc and 6 RRd stars. \citet{Kuehn2013} estimated an accuracy of $10^{-5}$ days for the periods of the RRab and RRc variables and an uncertainty approximately one order of magnitude larger for the periods of the RRd stars.

%Reticulum contains 32 RRLs for which accurate periods and photometric parameters in the $B$, $V$, $I$ and $K$ bands were obtained in previous studies (\citealt{Demers1976}, \citealt{Walker1992}, \citealt{DallOra2004}, \citealt{Kuehn2013}). 
All RRLs in the cluster share the same reddening, distance and metallicity  \citep{Kuehn2013}. Additionally,  the majority of the RRLs in the cluster are  located in uncrowded fields, which allows us to accurately determine their mean magnitudes in the 3.6 and 4.5 $\mu m$ passbands. 

 In this paper we present the first MIR $PL$ relations defined using RRLs in a system outside our Galaxy, and compare them with empirical studies of the MIR $PL$ relation based on field and GC MW RRLs  (\citealt{Madore2013}, \citealt{Dambis2014}, \citealt{Klein2014}, \citealt{Neeley2015}, \citealt{Clementini2017}).

This study is part of the Carnegie RR Lyrae Program (CRRP, PI Freedman). CRRP is a Warm {\it Spitzer} Cycle 9 Exploration Science program which uses the [3.6] and [4.5] channels of the Infrared Array Camera (IRAC, \citealt{Fazio2004}) to study MW RRL. In addition to the Reticulum observations described here, CRRP is also studying $\sim 30$ GCs in the MW and LMC, a sample of field RRLs in the MW \citep{Monson2017}, and a large number of bulge RRLs. CRRP is complementary program to the {\it Spitzer} Merger History and Shape of the Halo Program (SMHASH, PI Johnston), which uses halo, stream,  GC, and dwarf spheroidal galaxy (dSph) RRLs to examine the MW's structure (e.g. \citealt{Neeley2017}, \citealt{Hendel2017}, \citealt{Garofalo2018}).

%Warm Spitzer program in Cycle 9 using the . 

To calibrate the zero point of our MIR $PL$ relations  we used trigonometric parallaxes of five Galactic RRLs measured with the Fine Guide Sensor (FGS) on board of the {\it Hubble Space Telescope (HST)} by \citet{Benedict2011}. As a comparison, we have also derived the zero points for our MIR $PL$ relations using the trigonometric parallaxes for these same stars from the first data release (DR1) of the European Space Agency (ESA) mission {\it Gaia} (\citealt{Prusti2016}, \citealt{Brown2016}), that were computed as part of the  Tycho-{\it Gaia} Astrometric Solution (TGAS, \citealt{Lindegren2016}),  in addition to the most recent {\it Gaia} parallaxes released with the second Data Release (DR2, \citealt{Brown2018}; \citealt{Lindegren2018}).
% and, 
%as an alternative,  we used the trigonometric parallaxes published for the same stars in the first data release (DR1) of the European Space Agency (ESA) mission {\it Gaia} (\citealt{Prusti2016}, \citealt{Brown2016}), that were  computed as part of the Tycho-{\it Gaia} Astrometric Solution (TGAS, \citealt{Lindegren2016}), as well as {\bf the parallaxes released with the second {\it Gaia} Data Release (DR2, \citealt{Brown2018}; \citealt{Lindegren2018})}.

The paper is organised as follows. 
In Section~\ref{sec:data}, we describe the observations and data reduction, and provide  MIR mean magnitudes and amplitudes of the 30 RRLs in Reticulum that form the basis of the present study. In Section~\ref{sec:pl} we present the results of fitting the RRL $PL$ relations in the 3.6 and 4.5 $\mu m$ passbands and describe the calibration of their zero points. We also present a comparison of our $PL$ relations with those in the literature. In Section~\ref{sec:dist} we use our new  relations to estimate the  distance to Reticulum. A summary of our results is provided in Section~\ref{sec:sum}.

\section{Data}\label{sec:data}
\subsection{Observations and data reduction}\label{subsec:data}
\label{sec:dat}

%\begin{figure*}
%  %\includegraphics[width=16cm]{RetMAPinverse_RRL1.eps}
%    \includegraphics[width=16cm, trim=40 300 40 300]{RetMAPinverse_RRL1_b.pdf}
%  \caption{Coverage of the  3.6 $\mm$ (magenta box) and 4.5 $\mm$ (cyan box) images of Reticulum. The cluster  is located at the intersection of the two fields of view. Open red, blue and green circles indicate the Reticulum RRab, RRc and RRd stars, respectively.}
%  \label{fig:map}
%\end{figure*}

\begin{figure*}
    \includegraphics[width=16cm, trim=30 0 40 0]{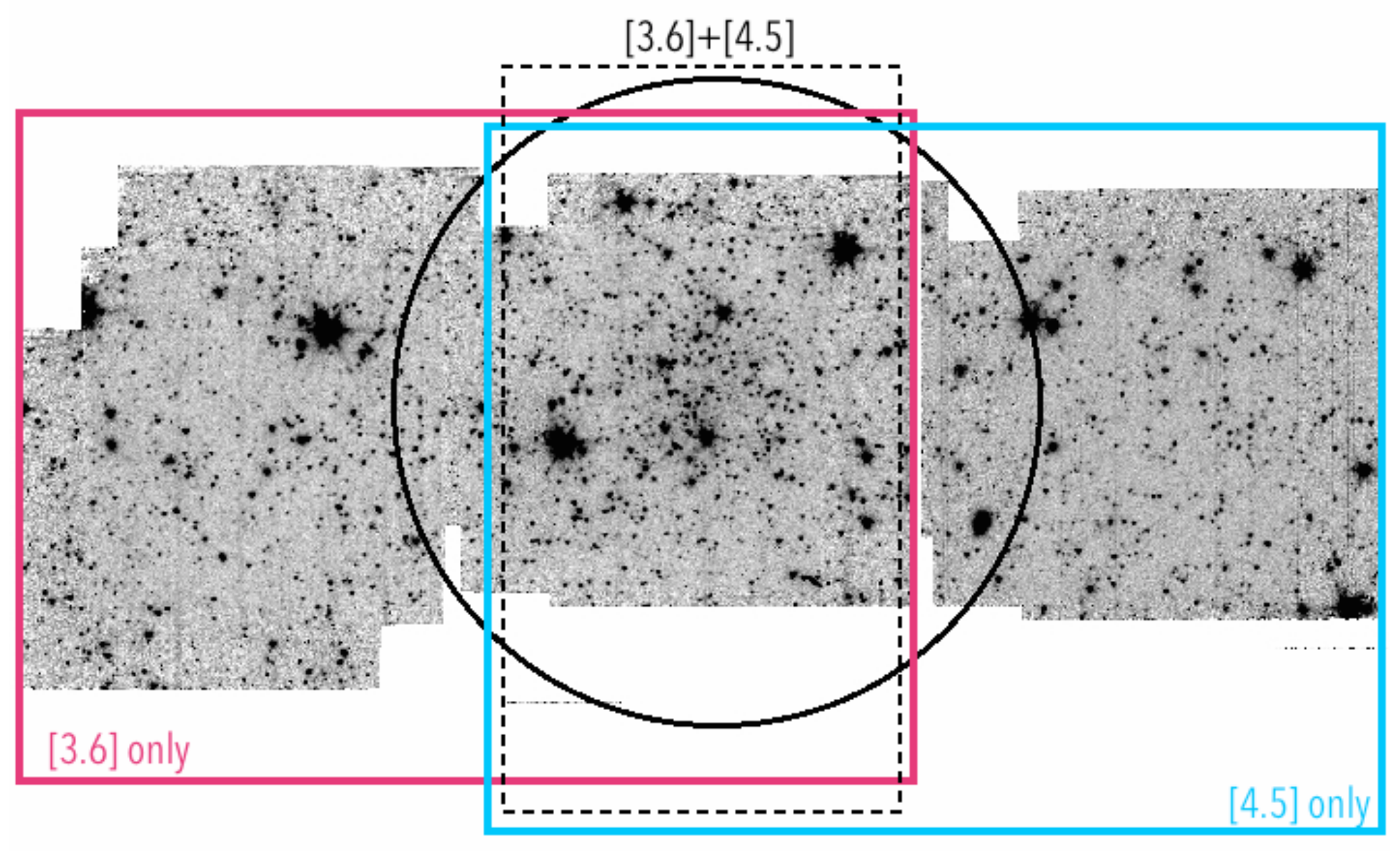}
  \caption{Coverage of the  3.6 $\mm$ (magenta box) and 4.5 $\mm$ (cyan box) images of Reticulum. The cluster  is located at the intersection of the two fields of view (black dashed line). The black circle shows the location of the cluster, and has a radius equal to two half-light radii of Reticulum (2.35~arcmin,  \citealt{Bica1999}).}
  \label{fig:map1}
\end{figure*}

\begin{figure}
       \includegraphics[trim=140 160 40 100,width=\linewidth]{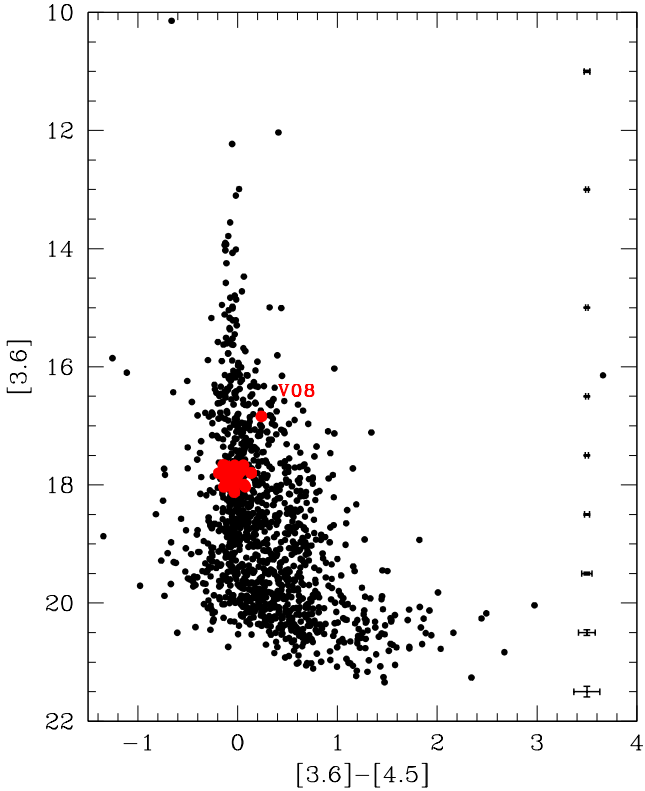}
  \caption{CMD ([3.6] vs [3.6]$-$[4.5])  of  1284 sources in the field of Reticulum observed with {\it Spitzer} in  both 3.6 and 4.5~$\mu m$ passbands. RRLs are marked in red.  Typical uncertainties of the photometry per magnitude bin estimated with ALLFRAME are shown on the right.}
  \label{fig:cmd_mi}
\end{figure}

\begin{figure*}
 \includegraphics[trim=0 140 0 120,width=18cm]{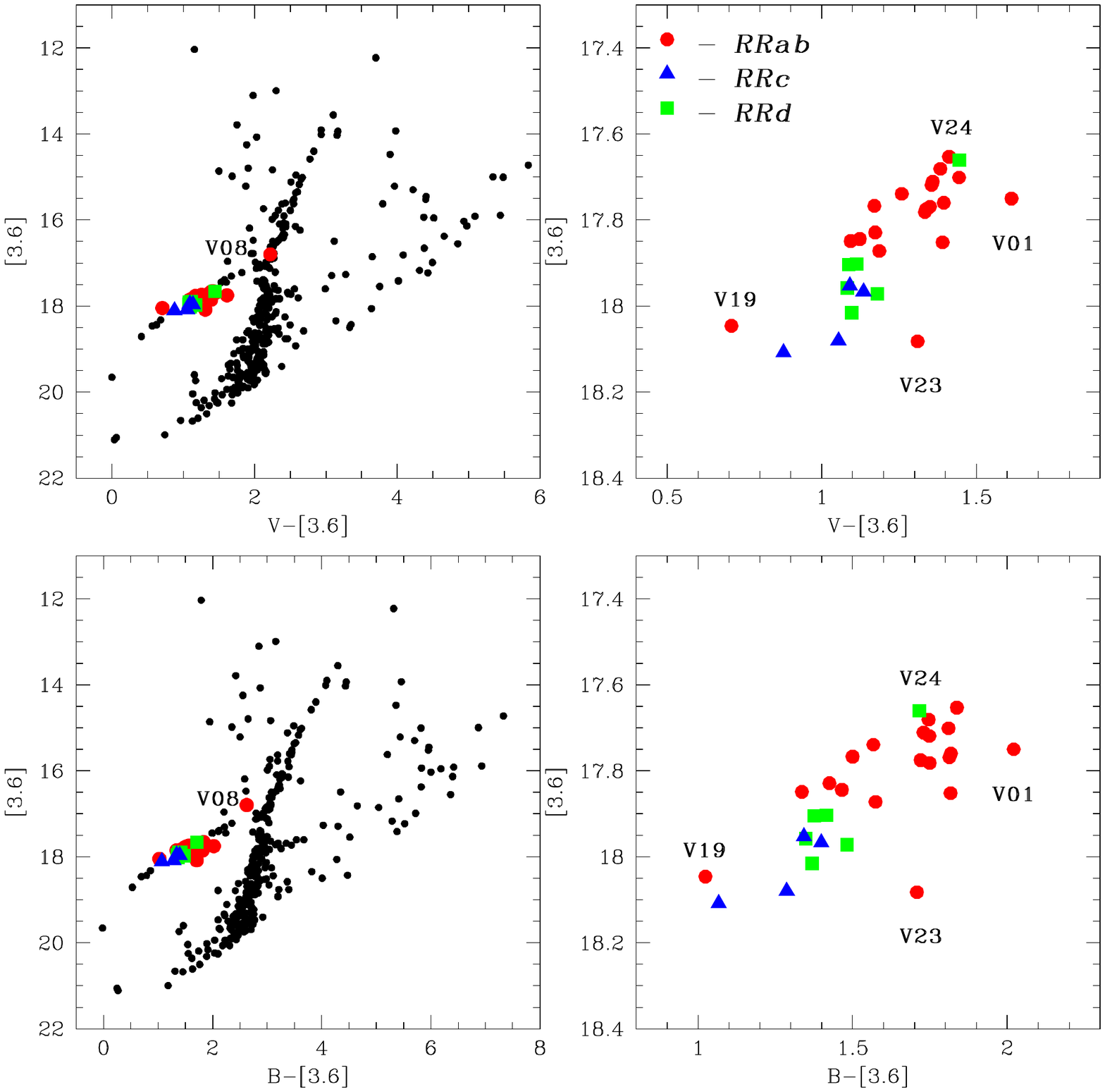}
  \caption{\textit{ Left panels:} CMDs ([3.6] {\it vs} $V -$ [3.6] and [3.6] {\it vs} $B -$ [3.6], in the upper- and bottom-left panels, respectively)  of  364 sources in the field of the Reticulum cluster that we observed with {\it Spitzer}.  $B$ and $V$ photometry for these objects is available from \citet{Jeon2014}. Red circles, blue triangles and green squares represent RRab, RRc and RRd stars, respectively. \textit{Right panels:} Zoom-in of the  region populated by RRLs in the CMDs shown in the left panels. For clarity reasons we do not plot the RRL V08  in the right panels of the figure.}
  \label{fig:cmd}
\end{figure*}

\begin{figure*}
      \includegraphics[width=16cm, trim=30 0 40 0]{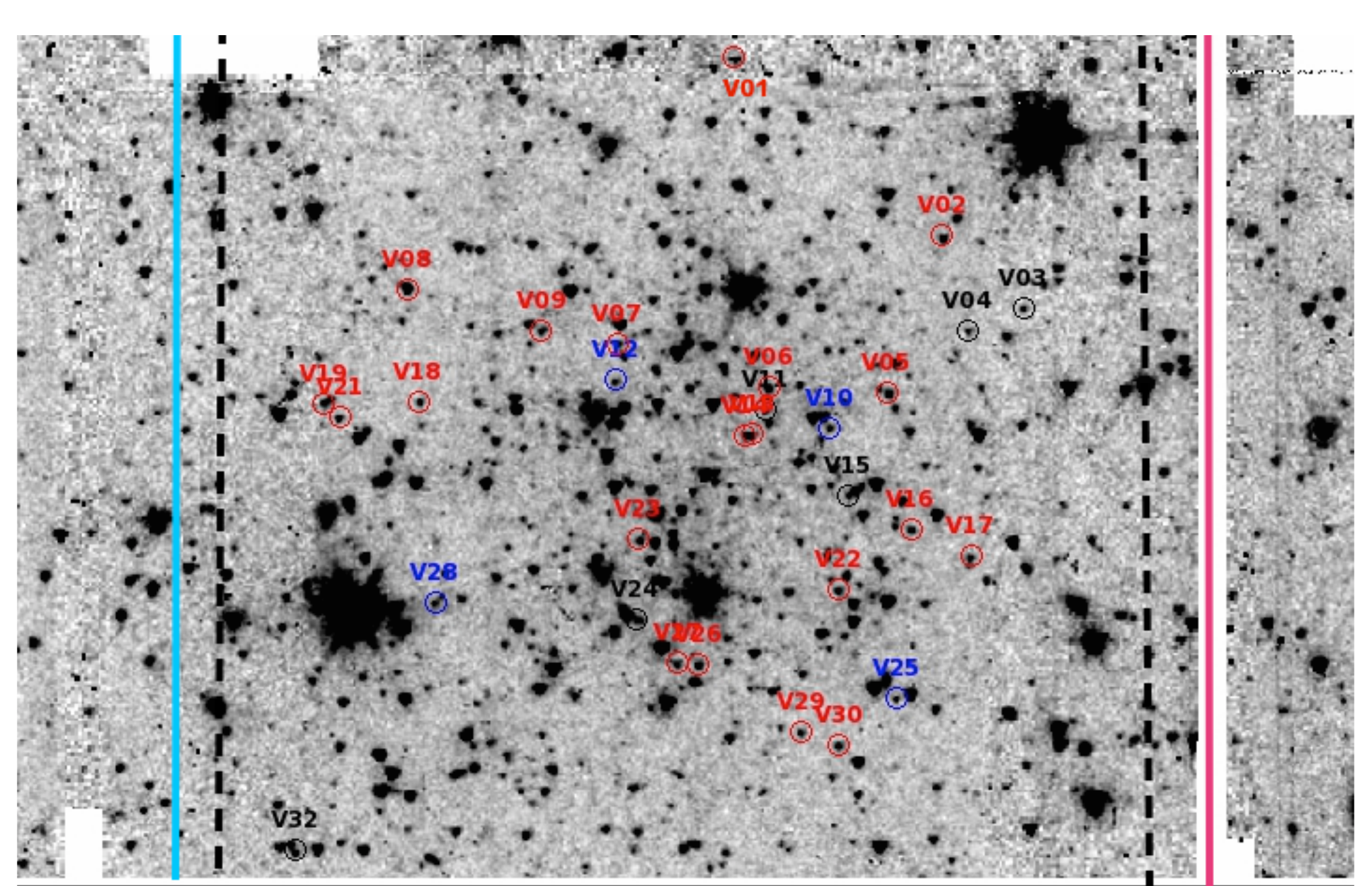}
  \caption{Zoom-in of Fig.~\ref{fig:map1} showing the region of the Reticulum cluster. Open red, blue and black circles indicate the Reticulum RRab, RRc and RRd stars, respectively.}
  \label{fig:map2}
\end{figure*}

The {\it Spitzer Space Telescope} \citep{Werner2004} was launched on 2003 August 25 with the main goal of  providing  a unique, infrared view of the universe. It operates at wavelengths which have low sensitivity to extinction and it has proven to be a perfect tool for undertaking variable star studies \citep{Freedman2011} as it 
%{\it Spitzer}  
was able to be scheduled so as to uniformly sample the light curves of variable stars providing high-precision stellar photometry that is very stable in its calibrated zero point.
%Moreover, it operates at wavelengths which have low sensitivity to extinction.

Reticulum was observed during Cycle 9 of the Warm {\it Spitzer } mission.  The observations were executed  on 2012  November 27 and consist of 12 epochs (each with individual exposures of $\sim31$~s), simultaneously in the 3.6 and 4.5 $\mu m$ passbands. The observations were distributed over a time interval of approximately 14 hours, corresponding to the longest period RRL in the field.

%The Reticulum cluster in the LMC was observed during the Warm {\it Spitzer } mission simultaneously in the 3.6 and 4.5 $\mu m$ passbands. The observations were executed  on 2012  November 27 and consist of 12 epochs (each with individual exposures of $\sim31$~s) in each Warm IRAC passband distributed over a time interval of approximately 14 hours, corresponding to the longest period RRL in the field. 
%Numbers of epochs and distribution of observations in time were optimised to allow a good coverage of the RRLs light curves. 
This spacing and time coverage were chosen to optimise the phase coverage of the RRLs that were known to exist in the field. 
The total area mapped by the {\it Spitzer} observations of Reticulum is shown in Figure~\ref{fig:map1}. The cluster is located at the intersection of the {\it Spitzer} [3.6] and [4.5] fields of view. The elongated field shape is owing to the dither pattern of the observations. While ``on target" observations are taken in one IRAC band, adjacent ``off target'' observations are taken simultaneously in the other passband. Colour information is available for the central region, and single band data  are available for the two side regions.
We outlined the cluster with a black circle of radius equal to two half-light radii of Reticulum (2.35~arcmin,  \citealt{Bica1999}).

\begin{table*}
\begin{minipage}{18cm}

 \caption{ Main properties of the 364 sources in common between our MIR catalogues and the visual catalogues of \citet{Jeon2014}}\label{tab:jeon}

\begin{tabular}{|l|c|c|c|c|c|c|c|c|}

\hline
\hline
ID Jeon$^{a}$ & ID Kuehn$^{b}$ & RA  & Dec & Flag$^{c}$  & B & V & [3.6]$^{d}$  &  [4.5]$^{d}$ \\ 
{}&{}&(J2000)&(J2000)& {}& (mag) & (mag) & (mag) & (mag)\\
\hline
Reticulum   225   &    V01  &  04:35:51.5 &  $-$58:51:03.4 &  1  &   19.771 $\pm$ 0.115  &  19.364 $\pm$ 0.084  &  17.751    &     17.826      \\  
Reticulum   279   &    V02  &  04:35:56.6 &  $-$58:52:32.1 &  1  &   19.467 $\pm$ 0.256  &  19.074 $\pm$ 0.205  &  17.717    &     17.678    \\       
Reticulum   269   &    V03  &  04:35:58.6 &  $-$58:53:06.7 &  1  &   19.385 $\pm$ 0.216  &  19.112 $\pm$ 0.163  &  17.997    &     17.932    \\      
Reticulum   272   &    V04  &  04:36:00.3 &  $-$58:52:50.0 &  1  &   19.307 $\pm$ 0.218  &  19.040 $\pm$ 0.178  &  17.963    &     17.974    \\       
Reticulum   250   &    V05  &  04:36:04.1 &  $-$58:52:29.3 &  1  &   19.440 $\pm$ 0.286  &  19.070 $\pm$ 0.239  &  17.669    &     17.611     \\       
Reticulum   294   &    V06  &  04:36:05.4 &  $-$58:51:47.8 &  1  &   19.495 $\pm$ 0.257  &  19.112 $\pm$ 0.216  &  17.747    &     17.826    \\       
Reticulum   220   &    V07  &  04:36:05.5 &  $-$58:50:51.8 &  1  &   19.185 $\pm$ 0.504  &  18.943 $\pm$ 0.421  &  17.821    &     17.978    \\       
Reticulum   288   &    V08  &  04:36:05.7 &  $-$58:49:34.3 &  1  &   19.421 $\pm$ 0.189  &  19.016 $\pm$ 0.129  &  16.841    &     16.602    \\      
Reticulum   237   &    V09  &  04:36:05.9 &  $-$58:50:24.5 &  1  &   19.306 $\pm$ 0.341  &  18.998 $\pm$ 0.286  &  17.726    &     17.756    \\       
Reticulum   273   &    V10  &  04:36:06.4 &  $-$58:52:12.6 &  1  &   19.365 $\pm$ 0.224  &  19.102 $\pm$ 0.200  &  17.960    &     18.002    \\       
Reticulum   259   &    V11  &  04:36:06.4 &  $-$58:51:48.7 &  1  &   19.280 $\pm$ 0.268  &  18.992 $\pm$ 0.228  &  17.913    &     18.037     \\       
Reticulum   243   &    V12  &  04:36:07.1 &  $-$58:50:54.9 &  1  &   19.175 $\pm$ 0.105  &  18.984 $\pm$ 0.084  &  18.124    &     18.157    \\       
Reticulum   255   &    V13  &  04:36:07.8 &  $-$58:51:46.8 &  1  &   19.531 $\pm$ 0.322  &  19.116 $\pm$ 0.255  &  17.773    &     17.743    \\       
Reticulum   244   &    V14  &  04:36:07.8 &  $-$58:51:44.4 &  1  &   19.427 $\pm$ 0.324  &  19.065 $\pm$ 0.246  &  17.683    &     17.792    \\       
Reticulum   262   &    V15  &  04:36:09.1 &  $-$58:52:25.8 &  1  &   19.453 $\pm$ 0.162  &  19.152 $\pm$ 0.142  &  17.980    &     18.050    \\       
Reticulum   233   &    V16  &  04:36:09.8 &  $-$58:52:50.7 &  1  &   19.254 $\pm$ 0.436  &  19.002 $\pm$ 0.340  &  17.814    &     17.823    \\       
Reticulum   298   &    V17  &  04:36:10.3 &  $-$58:53:13.6 &  1  &   19.267 $\pm$ 0.419  &  18.937 $\pm$ 0.313  &  17.754    &     17.850    \\       
Reticulum   261   &    V18  &  04:36:10.6 &  $-$58:49:50.2 &  1  &   19.511 $\pm$ 0.308  &  19.145 $\pm$ 0.263  &  17.733    &     17.760    \\       
Reticulum   227   &    V19  &  04:36:11.9 &  $-$58:49:18.1 &  1  &   19.070 $\pm$ 0.449  &  18.754 $\pm$ 0.332  &  18.029    &     17.952    \\       
Reticulum   287   &    V21  &  04:36:12.3 &  $-$58:49:24.3 &  1  &   19.578 $\pm$ 0.283  &  19.155 $\pm$ 0.217  &  17.798    &     17.663     \\      
Reticulum   276   &    V22  &  04:36:13.4 &  $-$58:52:32.0 &  1  &   19.309 $\pm$ 0.363  &  18.967 $\pm$ 0.260  &  17.801    &     17.989    \\      
Reticulum   258   &    V23  &  04:36:13.8 &  $-$58:51:19.0 &  1  &   19.790 $\pm$ 0.134  &  19.392 $\pm$ 0.120  &  18.034    &     18.170    \\      
Reticulum   248   &    V24  &  04:36:17.3 &  $-$58:51:26.6 &  1  &   19.377 $\pm$ 0.273  &  19.106 $\pm$ 0.229  &  17.660    &     17.803    \\      
Reticulum   277   &    V25  &  04:36:17.4 &  $-$58:53:02.6 &  1  &   19.367 $\pm$ 0.255  &  19.134 $\pm$ 0.178  &  18.058    &     18.097    \\      
Reticulum   296   &    V26  &  04:36:18.5 &  $-$58:51:51.8 &  1  &   19.490 $\pm$ 0.155  &  19.065 $\pm$ 0.113  &  17.668    &     17.698    \\      
Reticulum   295   &    V27  &  04:36:18.7 &  $-$58:51:44.3 &  1  &   19.669 $\pm$ 0.263  &  19.243 $\pm$ 0.200  &  17.779    &     17.837    \\      
Reticulum   246   &    V28  &  04:36:19.2 &  $-$58:50:15.5 &  1  &   19.294 $\pm$ 0.223  &  19.043 $\pm$ 0.166  &  18.003    &     18.071    \\      
Reticulum   223   &    V29  &  04:36:20.1 &  $-$58:52:33.5 &  1  &   19.446 $\pm$ 0.381  &  19.058 $\pm$ 0.285  &  17.832    &     17.847     \\      
Reticulum   236   &    V30  &  04:36:20.2 &  $-$58:52:47.7 &  1  &   19.582 $\pm$ 0.268  &  19.119 $\pm$ 0.284  &  17.837    &     17.835    \\     
Reticulum   268   &    V32  &  04:36:32.0 &  $-$58:49:53.2 &  1  &   19.319 $\pm$ 0.207  &  19.016 $\pm$ 0.160  &  17.953    &     18.003    \\       
Reticulum   5     &    $-$  &  04:35:52.8 &  $-$58:49:40.5 &  0  &   13.825 $\pm$ 0.017  &  13.191 $\pm$ 0.020  &  12.034  &     11.625      \\        
Reticulum   12    &    $-$  &  04:35:50.9 &  $-$58:52:55.7 &  0  &   15.661 $\pm$ 0.014  &  14.151 $\pm$ 0.014  &  10.144   &     10.805      \\        
Reticulum   19    &    $-$  &  04:36:19.4 &  $-$58:54:20.6 &  0  &   15.950 $\pm$ 0.006  &  15.081 $\pm$ 0.016  &  13.103   &     13.121      \\        
Reticulum   23    &    $-$  &  04:36:12.3 &  $-$58:54:04.4 &  0  &   16.149 $\pm$ 0.009  &  15.294 $\pm$ 0.017  &  12.992   &     12.977      \\        
Reticulum   26    &    $-$  &  04:36:22.5 &  $-$58:49:08.2 &  0  &   16.212 $\pm$ 0.008  &  15.541 $\pm$ 0.016  &  13.787   &     13.880      \\    
 \hline

\end{tabular}
\medskip

$^{a}$ Identification from \citet{Jeon2014}.\\
$^{b}$ Identification from \citet{Kuehn2013}.\\
$^{c}$ A ``1" flag indicates objects from \citet{Jeon2014} that are classified as RRLs by \citet{Kuehn2013}, while a ``0" flag indicates  stars that do not have a counterpart in the \citet{Kuehn2013} catalogue of RRLs.\\
$^{d}$  Mean magnitudes in the {\it Spitzer} passbands obtained as a result of the data reduction procedure; uncertainties in magnitudes are not provided, since  \texttt{ALLFRAME} is known to provide underestimated uncertainties.\\

This table is published in its entirety at the CDS;  a portion is shown here for guidance with respect to its form and content.

\end{minipage}
\end{table*}

For each epoch a single epoch mosaic image was created from the individual basic calibrated data (BCD) frames provided by the {\it Spitzer} Science Center, using MOPEX \citep{MOPEX}. Mosaicked location-correction images for each epoch were also created. 
Point-spread function (PSF) photometry of the mosaicked images was performed using the \texttt{DAOPHOT-ALLSTAR-ALLFRAME} packages (\citealt{Stetson1987,Stetson1994}). Firstly, each image mosaic was converted from $MJy/sr$ to counts (DN) using the conversion factor  $(MJy/sr$ per $DN/sec)$ and the exposure time provided in the image header.
On each epoch the PSF model was created selecting $\sim$ 50  stars uniformly distributed across the frames ( with 30\% of stars in common among all epochs). Reticulum has an uncrowded nature, our PSF stars are bright  ($m_{[3.6]}<16.741$ mag and $m_{[4.5]}<16.519$ mag) and well isolated. With the \texttt{ALLSTAR} algorithm we fitted the PSF model given by the PSF stars to the sources in each frame. The \texttt{ALLFRAME} routine was then used to perform the PSF photometry  simultaneously on images at all epochs. The  \texttt{ALLFRAME} catalogue provides instrumental magnitudes for each star in the images with an arbitrary zero point value. %and does not take into account exposure time, extinction, or colour. 

To transform the photometry to standard IRAC Vega magnitudes we selected 53 stars in the 3.6 $\mm$ and 55 stars in the 4.5 $\mu m$ images,  homogeneously distributed over the frames. We obtained aperture photometry of these stars, calibrated to the zero point magnitude (zmag) provided in the IRAC handbook\footnote{\url{http://irsa.ipac.caltech.edu/data/SPITZER/docs/irac/iracinstrumenthandbook}, Chapter 4} which was then compared with the  \texttt{DAOPHOT} instrumental magnitudes in order to derive the final calibrated magnitudes. Point source photometry requires additional array location-dependent photometric corrections. These corrections improve the photometric accuracy for IRAC data by taking into account scattering and distortions, but above all, variations in the effective central wavelengths of the IRAC filters, which depend on the angle of incidence. 
%that therefore are position dependent of each source on the array .
%We performed separate data reductions for the two filters. 
%We obtained 3.6 $\mu m$ and 4.5 $\mu m$ photometric catalogues containing 4068 and 3379 sources, respectively. The 3.6 $\mu m$ magnitudes of the sources in our catalogue range from 21.57 to 9.23 mag and have mean error of $\sim$ 0.04 mag, while the sources in the 4.5 $\mu m$ have magnitudes from 21.11 to 10.36 mag and mean error of $\sim$ 0.04 mag.
The final photometric catalogues contain 4068 and 3379 sources in 3.6 and 4.5 $\mu m$, respectively. The 3.6 $\mu m$  photometry spans the range from  21.57 to 9.23~mag, while the 4.5 $\mu m$  photometry ranges from 21.11 to 10.36 mag,  where upper limits are owing to saturation.  Uncertainties  in the 3.6 $\mu m$ magnitudes range from 0.003 to 0.100~mag, while uncertainties in the 4.5 $\mu m$ magnitudes span the range from 0.005 to 0.100~mag. The typical uncertainties at the level of the Reticulum HB ($\sim18$~mag) are 0.017~mag  and 0.022~mag in the 3.6 $\mu m$ and 4.5 $\mu m$ bands, respectively. Typical uncertainties of the photometry per magnitude bin are shown in Fig.~\ref{fig:cmd_mi}. We remind the reader that the uncertainties in magnitudes calculated as part of the data reduction procedure using \texttt{ALLFRAME} software can be significantly underestimated.

%from  at the lower magnitudes are  Both bands have a mean photometric uncertainty of 0.04 mag.
We used  \texttt{DAOMASTER} and  \texttt{DAOMATCH} to combine the 3.6 $\mm$ and 4.5 $\mm$ catalogues and obtained a list of 1284 sources for which photometry is available in both passbands. 
%The relatively small fraction of stars covered by both bands is expected; this is owing to the dither pattern of the observations (see Fig.~\ref{fig:map1}). 
The [3.6] vs $([3.6]-[4.5])$ CMD of these sources is shown in Fig.~\ref{fig:cmd_mi}, where red dots represent the RRLs. The MIR CMD has its limitations owing to low sensitivity of the luminosity in the MIR passbands to the temperature. 
%Indeed, the cluster's Red Giant Branch (RGB) and the HB are not well pronounced, even though RRLs in Fig.~\ref{fig:cmd_mi} form a well-defined population, both in colour and in magnitude.
 Although the  HB is not very well distinguished from the Red Giant Branch (RGB)   in these  passbands, the RRLs in Fig.~\ref{fig:cmd_mi}  tracing  a well-defined population, both in colour and magnitude  allow us to clearly identify the location of the cluster HB.

 \citet{Jeon2014} provide $B$, $V$ photometry for 766 sources in Reticulum.  We crossmatched the catalogue of \citet{Jeon2014} with our MIR catalogue of 1284 sources and found 364 objects in common. Information on these sources with the apparent $B$ and $V$ magnitudes provided by \citet{Jeon2014} and MIR magnitudes obtained here is presented in Table~\ref{tab:jeon}.
 %as a result of the data reduction procedure described above, is presented in Table~\ref{tab:jeon}.
 The [3.6], $(V-[3.6])$ and [3.6], $(B-[3.6])$ CMDs of  these 364 sources are shown in the left panels of Fig.~\ref{fig:cmd}. The cluster RGB and the HB are well defined, demonstrating the quality of our photometry and the accuracy of the calibration procedure. 
% Figure~\ref{fig:cmd} shows the colour-magnitude diagram (CMD) of 1284 sources which are in common between 3.6 $\mm$ and 4.5 $\mm$ fields of view.

\subsection{RRLs in Reticulum}\label{subsec:rr}

\begin{table*}
\tiny
\begin{minipage}{18cm}

 \caption{Identification and properties of the 30 RRLs in Reticulum for which we analysed the  {\it Spitzer} 3.6 and 4.5 $\mm$ time-series data.}\label{tab:gen}

\begin{tabular}{|l|c|c|l|c|c|c|c|c|c|c|r|c|}

\hline\hline
ID  & RA    & DEC   & Type &P  & Log(P) &  Epoch$^{a}$ & [3.6]$^{b}$   & ${\rm Amp_{[3.6]}}$ & [4.5]$^{b}$   & ${\rm Amp_{[4.5]}}$& $\varpi$~~~~~~~ & Flag$^{c}$  \\ 
{}&(J2000)&(J2000)& {}& (days) & {} &  (HJD) & {(mag)}& {(mag)} & (mag)& (mag)& (mas)~~~~~& {}\\
\hline

  V01    & 04:35:51.5 & $-$58:51:03.4 & RRab  &  0.50993 & $-$0.29249 & 56258.032 & $17.750\pm0.109$ & 0.410 & $17.825\pm0.185$ & 0.217 & $-0.18\pm0.19$&1\\       
  V02    & 04:35:56.6 & $-$58:52:32.1 & RRab  &  0.61869 & $-$0.20853 & 56257.239 & $17.719\pm0.047$ & 0.153 & $17.603\pm0.108$ & 0.238 & $ 0.13\pm0.18$&0\\       
  V03    & 04:35:58.6 & $-$58:53:06.7 & RRd   &  0.35350 & $-$0.45161 & 56257.592 & $18.015\pm0.068$ & 0.204 & $17.918\pm0.077$ & 0.244 & $-0.02\pm0.18$&0\\       
  V04    & 04:36:00.3 & $-$58:52:50.0 & RRd   &  0.35320 & $-$0.45198 & 56268.034 & $ 17.958\pm0.056$ & $-$ & $17.908\pm0.103$ & $-$ & $ 0.20\pm0.18$&0\\       
  V05    & 04:36:04.1 & $-$58:52:29.3 & RRab  &  0.57185 & $-$0.24272 & 56257.444 & $17.711\pm0.042$ & 0.249 & $17.617\pm0.063$ & 0.322 & $ 0.14\pm0.21$&0\\       
  V06-BL & 04:36:05.4 & $-$58:51:47.8 & RRab  &  0.59526 & $-$0.22529 & 56257.439 & $17.775\pm0.041$ & 0.276 & $17.862\pm0.094$ & 0.363 & $-0.01\pm0.18$&0\\       
  V07    & 04:36:05.5 & $-$58:50:51.8 & RRab  &  0.51044 & $-$0.29206 & 56257.550 & $17.849\pm0.052$ & 0.280 & $18.010\pm0.097$ & 0.255 & $ 0.43\pm0.23$&0\\       
  V08    & 04:36:05.7 & $-$58:49:34.3 & RRab  &  0.64496 & $-$0.19047 & 56257.608 & $16.796\pm0.055$ & 0.144 & $16.558\pm0.097$ & $-$ & $ 0.19\pm0.20$&1\\       
  V09    & 04:36:05.9 & $-$58:50:24.5 & RRab  &  0.54496 & $-$0.26364 & 56257.824 & $17.739\pm0.047$ & 0.166 & $17.732\pm0.069$ & 0.261 & $-0.09\pm0.17$&0\\       
  V10    & 04:36:06.4 & $-$58:52:12.6 & RRc   &  0.35256 & $-$0.45277 & 56257.813 & $17.967\pm0.044$ & 0.185 & $17.964\pm0.077$ & 0.166 & $ 0.02\pm0.18$&0\\       
  V11    & 04:36:06.4 & $-$58:51:48.7 & RRd   &  0.35540 & $-$0.44928 & 56257.987 & $17.904\pm0.045$ & 0.160 & $18.000\pm0.103$ & 0.160 & $ 0.02\pm0.21$&0\\       
  V12    & 04:36:07.1 & $-$58:50:54.9 & RRc   &  0.29627 & $-$0.52831 & 56258.029 & $ 18.108\pm0.056$ & $-$ & $18.184\pm0.104$ & $-$ & $ 0.11\pm0.19$&0\\       
  V13    & 04:36:07.8 & $-$58:51:46.8 & RRab  &  0.60958 & $-$0.21497 & 56257.456 & $17.782\pm0.039$ & 0.175 & $17.716\pm0.055$ & 0.209 & $-0.05\pm0.18$&0\\       
  V14-BL &  04:36:07.8 & $-$58:51:44.4 & RRab  &  0.58661 & $-$0.23165 & 56257.679 & $17.681\pm0.044$ & 0.288 & $17.728\pm0.052$ & 0.289 & $ 0.10\pm0.19$&0\\       
  V15    & 04:36:09.1 & $-$58:52:25.8 & RRd   &  0.35430 & $-$0.45063 & 56257.790 & $17.972\pm0.053$ & 0.138 & $18.040\pm0.115$ & 0.200 & $-0.04\pm0.20$&0\\       
  V16    & 04:36:09.8 & $-$58:52:50.7 & RRab  &  0.52290 & $-$0.28158 & 56257.913 & $17.829\pm0.056$ & 0.246 & $17.827\pm0.073$ & 0.210 & $ 0.10\pm0.19$&0\\       
  V17    & 04:36:10.3 & $-$58:53:13.6 & RRab  &  0.51241 & $-$0.29038 & 56257.599 & $17.767\pm0.045$ & 0.302 & $17.757\pm0.076$ & 0.340 & $ 0.39\pm0.20$&0\\       
  V18    & 04:36:10.6 & $-$58:49:50.2 & RRab  &  0.56005 & $-$0.25177 & 56257.751 & $17.701\pm0.050$ & 0.291 & $17.722\pm0.066$ & 0.247 & $-0.50\pm0.19$&0\\       
  V19    & 04:36:11.9 & $-$58:49:18.1 & RRab  &  0.48485 & $-$0.31439 & 56257.635 & $18.046\pm0.074$ & 0.360 & $17.981\pm0.155$ & 0.603 & $ 0.30\pm0.20$&1\\       
  V21    & 04:36:12.3 & $-$58:49:24.3 & RRab  &  0.60700 & $-$0.21681 & 56257.933 & $17.760\pm0.065$ & 0.292 & $17.665\pm0.089$ & 0.263 & $-0.05\pm0.20$&0\\       
  V22    & 04:36:13.4 & $-$58:52:32.0 & RRab  &  0.51359 & $-$0.28938 & 56257.851 & $17.844\pm0.072$ & 0.291 & $17.880\pm0.077$ & 0.351 & $-0.35\pm0.22$&0\\       
  V23-BL & 04:36:13.8 & $-$58:51:19.0 & RRab  &  0.46863 & $-$0.32917 & 56258.153 & $18.082\pm0.070$ & 0.244 & $18.245\pm0.114$ & 0.366 & $-0.01\pm0.21$& 0/1\\       
  V24    & 04:36:17.3 & $-$58:51:26.6 & RRd   &  0.34750 & $-$0.45905 & 56257.864 & $17.661\pm0.067$ & 0.089 & $17.780\pm0.119$ & $-$ & $-0.12\pm0.19$&1\\       
  V25    & 04:36:17.4 & $-$58:53:02.6 & RRc   &  0.32991 & $-$0.48160 & 56258.020 & $18.080\pm0.065$ & 0.191 & $18.073\pm0.077$ & 0.242 & $ 0.11\pm0.19$&0\\       
  V26    & 04:36:18.5 & $-$58:51:51.8 & RRab  &  0.65696 & $-$0.18246 & 56257.940 & $17.653\pm0.042$ & 0.100 & $17.720\pm0.087$ & 0.222 & $-0.11\pm0.19$&0\\       
  V27    & 04:36:18.7 & $-$58:51:44.3 & RRab  &  0.51382 & $-$0.28919 & 56257.982 & $17.852\pm0.052$ & 0.350 & $17.900\pm0.071$ & 0.331 & $-0.31\pm0.22$&0\\       
  V28    & 04:36:19.2 & $-$58:50:15.5 & RRc   &  0.31994 & $-$0.49493 & 56257.701 & $17.952\pm0.065$ & 0.163 & $18.056\pm0.141$ & $-$ & $ 0.23\pm0.22$&1\\       
  V29    & 04:36:20.1 & $-$58:52:33.5 & RRab  &  0.50815 & $-$0.29401 & 56257.682 & $17.872\pm0.067$ & 0.348 & $17.818\pm0.078$ & 0.440 & $-0.14\pm0.20$&0\\       
  V30    & 04:36:20.2 & $-$58:52:47.7 & RRab  &  0.53501 & $-$0.27164 & 56257.725 & $17.769\pm0.047$ & 0.242 & $17.750\pm0.076$ & 0.301 & $-0.19\pm0.18$&0\\       
  V32    & 04:36:32.0 & $-$58:49:53.2 & RRd   &  0.35230 & $-$0.45309 & 56257.750 & $17.903\pm0.060$ & 0.060 & $18.083\pm0.208$ & $-$ & $ 0.07\pm0.22$&1\\

\hline

\end{tabular}
\normalsize
\medskip

$^{a}$ Epoch of maximum light (HJD-2400000) of the 3.6 $\mm$ light curve. The epoch of maximum light is the same in the 3.6 $\mm$ and 4.5 $\mm$ passbands for all sources except for V10.
The 4.5 $\mm$ data of this star had to be shifted by 0.4 in phase. \\
$^{b}$ Magnitudes in the {\it Spitzer} passbands obtained by fitting a  model to the data with the GRATIS software.\\
$^{c}$ A ``0" in column 13 flags  objects which were used to fit the $PL$ relations whereas a ``1" flags stars that were discarded.  A ``0/1" flags V23 because the star was used to fit the the 3.6 $\mm$ $PL$ relation, but was automatically rejected by the 3-sigma clipping procedure when fitting the 4.5 $\mm$ $PL$ relation. See text for details. \\

\end{minipage}
\end{table*}

In this paper we study the MIR (3.6 and 4.5 $\mm$) time-series photometry of the Reticulum RRLs. We cross-matched our catalogue of 1284 sources with 3.6 and 4.5 $\mm$ photometry (see Section~\ref{subsec:data}) against 32 RRLs from \citet{Kuehn2013} catalogue.  Thirty RRLs from the \citeauthor{Kuehn2013} sample were identified, while we were not able to find the counterparts of V20 and V31 in our catalogue since these objects belong to non-resolved elongated groups of  sources in our data. The 30 RRLs  (20 RRab, 4 RRc and 6 RRd stars) are listed in Table~\ref{tab:gen} adopting the identification number, coordinates, period, and  classification in type of \citet{Kuehn2013}. Periods range from 0.29627 to 0.65696 days.   For the 6 RRd stars we list only the first-overtone periodicity. 
%MIR magnitudes for V20 and V31 were not reliably measured  since these two stars are blended with brighter companions in our Spitzer data. 
%We adopted the coordinates, periods, classification and identification numbers from \citet{Kuehn2013}.  They are listed in Table~\ref{tab:gen}. 
The positions of the 30 RRLs in the {\it Spitzer} images is shown in Fig.~\ref{fig:map2} with the RRab, RRc and RRd stars marked by red, blue and black open circles, respectively.  
The 3.6 and 4.5 $\mm$ light curves are presented in Fig.~\ref{fig:lc1}.
%The  30 RRLs  were observed in 12 epochs over $\sim$ 14 hours in both IRAC passbands that allowed a good coverage of the light curves which  are presented in Fig.~\ref{fig:lc1}.  

\begin{figure*}
  \includegraphics[width=16cm]{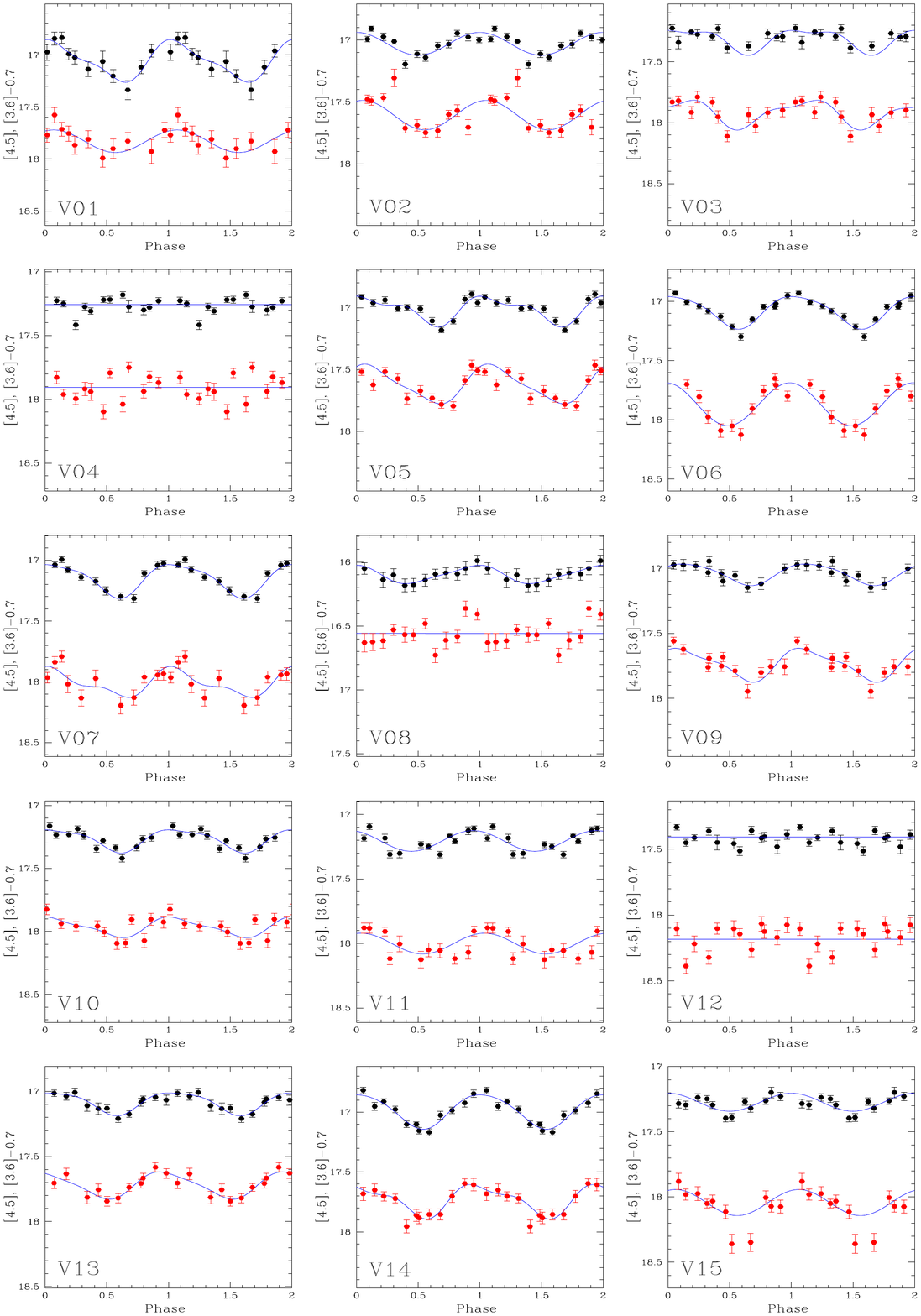}
  \caption{{\it Spitzer} MIR light curves of the RRLs in Reticulum analysed in the present study. Black and red dots represent the 3.6 $\mm$ and 4.5 $\mm$ data points, respectively,  with the 3.6 $\mm$ shifted by $-0.7$~mag, for clarity. Blue solid lines show the best fit models.  Only data points used in the model fitting are shown.}
  \label{fig:lc1}
\end{figure*}

\begin{figure*}
  \includegraphics[width=16cm]{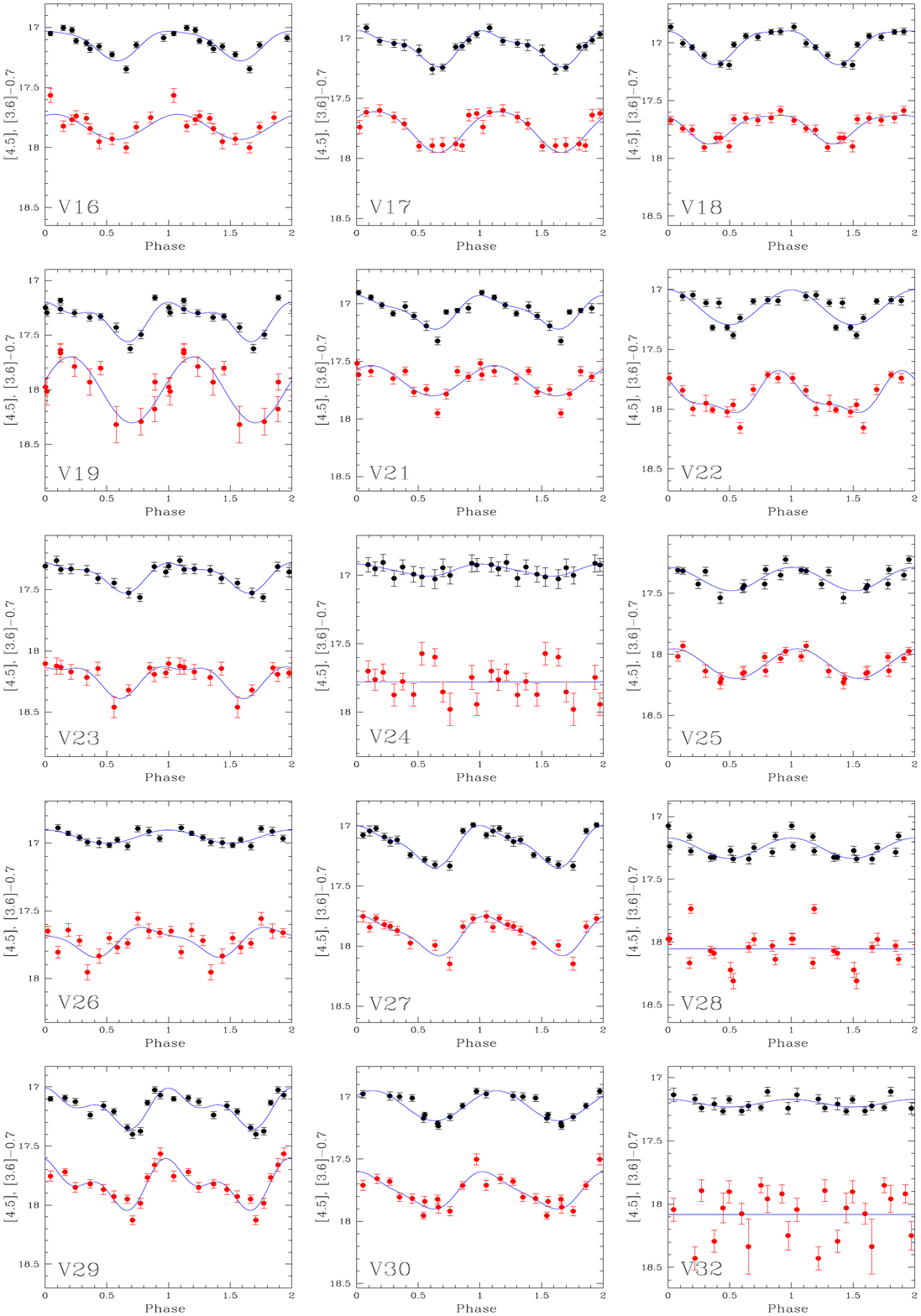}
  \contcaption{}
  %\label{fig:lc2}
\end{figure*}

The epoch of maximum light (Column~7 in Table~\ref{tab:gen}) is the same for the  3.6 and 4.5 $\mm$ passbands for all sources except for V10. The 4.5 $\mm$ data points of this star had to be shifted by $\sim$ 0.4 in phase when folded according to the   epoch in the 3.6 $\mm$ passband. 
 For a large majority of the RRLs we determined mean 3.6 and 4.5 $\mm$ magnitudes by Fourier fitting the light curves with the GRaphical Analyzer of TImes Series package (GRATIS, custom software  developed at the Observatory of Bologna by P. Montegriffo, see e.g. \citealt{Clementini2000}). The blue lines in Fig.~\ref{fig:lc1} show the best fit models obtained with GRATIS, while dots represent the data points retained in the analysis. We discarded obvious outliers,  however after  the cleaning procedure each source still has at least 11 data points,  since model fitting with the GRATIS software needs a minimum number of 11 data points to converge. % Mean magnitudes were derived as intensity averaged mean, that is by transforming to intensity each individual magnitude, then averaging the intensities and converting the mean intensity back into a magnitude. 
Mean magnitudes listed in Table~\ref{tab:gen} are intensity averages. 
The  standard deviation of the phase points from the fit was calculated considering only data points retained after removing outliers. The final uncertainty in the mean magnitude was calculated as the  sum in quadrature of the fit uncertainty provided by GRATIS and the photometric uncertainty ($\sqrt{\sum\sigma_{i}^2}/N$), where $\sigma_{i}$ is the photometric uncertainty of the individual observations and $N$ is the number of observations. Peak-to-peak amplitudes were also measured from the modelled light curves.

 For very noisy light curves, such as those of V04 and V12 in both the 3.6 and 4.5 $\mm$ passbands; and V08, V24, V28, V32 in the 4.5 $\mm$ passband, GRATIS failed to fit a reliable model to the data. For these stars the mean magnitudes were calculated as the weighted mean of the intensity values of each data point then transformed back to  magnitudes, while the uncertainty of the mean values was estimated as the standard deviation of the weighted mean. No amplitudes were calculated for these stars.

\begin{figure}
 \includegraphics[trim=20 120 0 100, width=\linewidth]{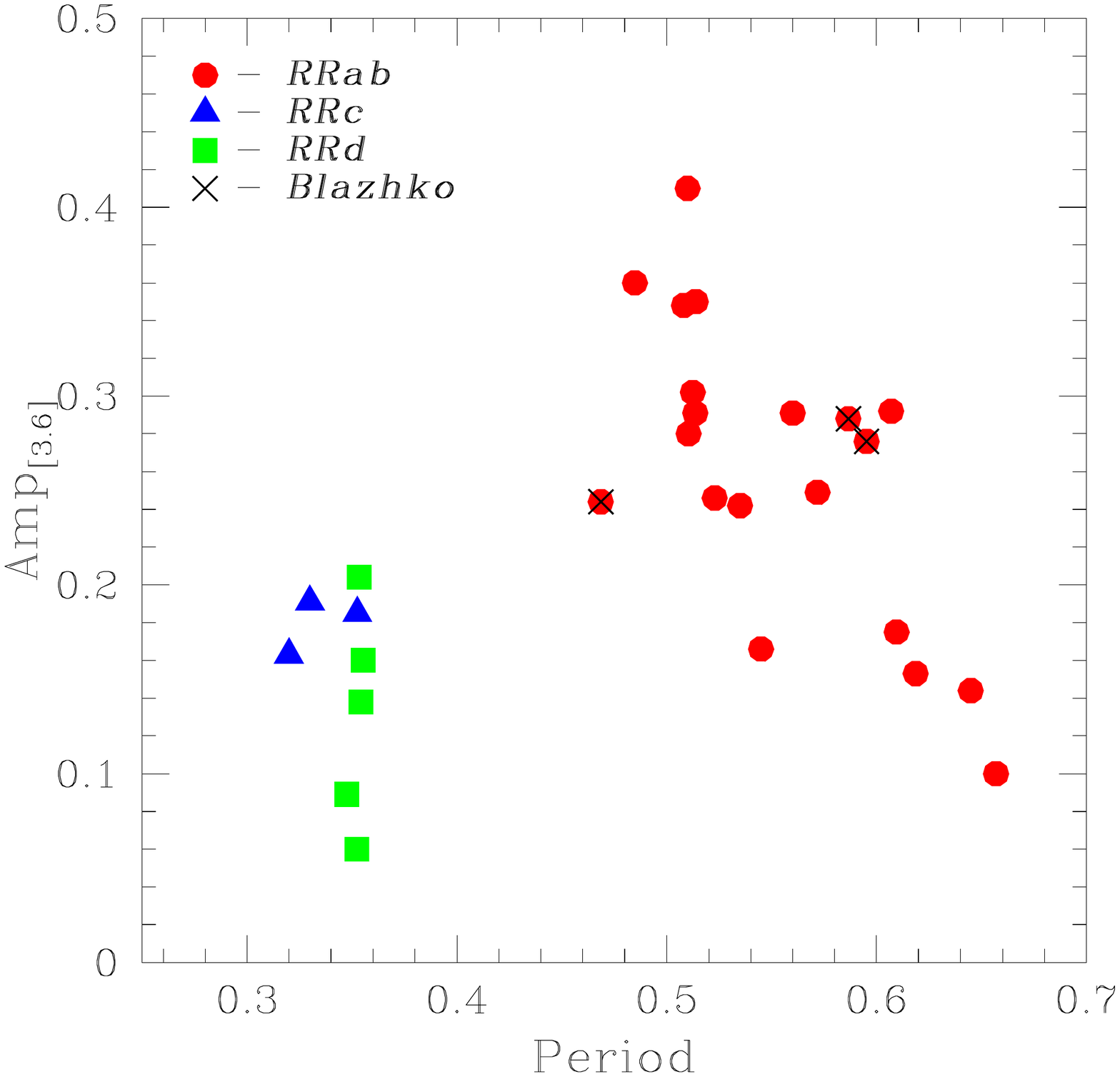}
 \includegraphics[trim=20 150 0 100, width=\linewidth]{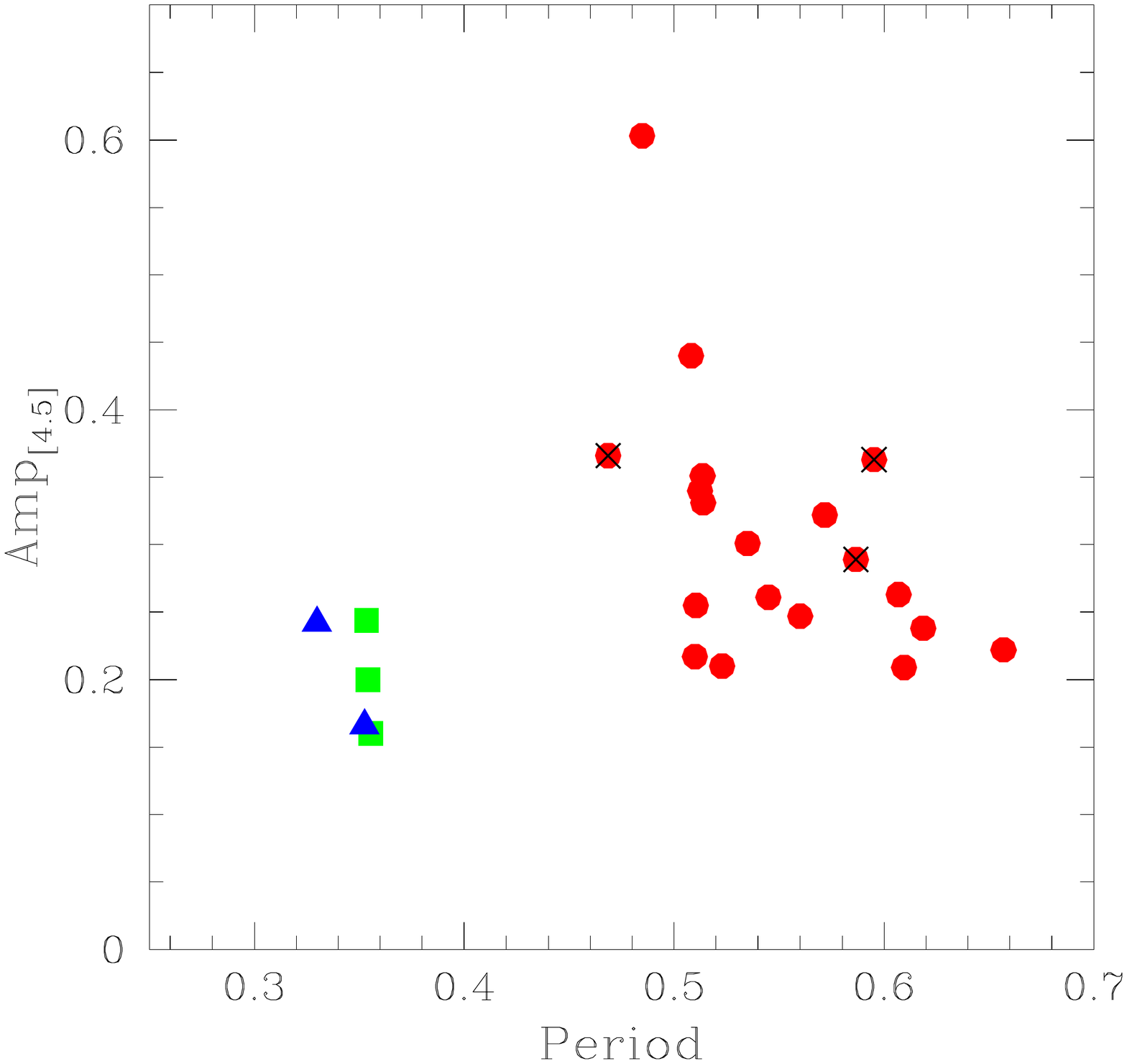}
  \caption{Period-amplitude (Bailey) diagrams in the 3.6 (upper panel) and 4.5 $\mm$ (lower panel) passbands defined  respectively by 28 and 24  RRLs in our sample,  for which the determination of the amplitude with GRATIS was possible. Red circles, blue triangles and green squares represent RRab, RRc and RRd stars, respectively. Stars  possibly affected by the Blazhko effect are marked by black crosses. }
  \label{fig:ampl}
\end{figure}

Columns from 8 to 11 in Table~\ref{tab:gen} provide  mean magnitudes with related uncertainties and amplitudes in the 3.6 and 4.5 $\mm$ bands for the RRLs in our sample, when available.
In  Fig.~\ref{fig:ampl} the amplitudes in the 3.6 and 4.5 $\mm$ bands are plotted versus period (Bailey diagrams)  for the samples of 28 and 24 RRLs, respectively, for which a determination of the amplitude from the light curve with GRATIS was possible.  RRab, RRc and RRd stars are shown by red circles, blue triangles and green squares, respectively. As expected, the  RRab stars are well  separated from  RRc and  RRd stars. The latter are located on the longest tail of the 
RRc period distribution. Consistently with the behaviour in the optical bands, the amplitude of the RRab stars decreases with increasing the period.
%RRc period distribution and in analogy with the behaviour at optical bands the amplitude of the RRab stars decreases with increasing the period. 
Black crosses in Fig.~\ref{fig:ampl} indicate three variables, V06, V14 and V23 that 
according to \citealt{Kuehn2013} possibly exhibit the Blazhko effect \citep{Blazko1907}, a modulation of both amplitude and shape of the light curve that typically  occurs on time spans ranging from a  few days to a few hundreds of days.  These stars are marked with a ``BL" flag in Table~\ref{tab:gen}. The  Blazhko effect  should not affect significantly the mean magnitude we measured   for the Blazhko  RRLs because the time interval covered by  our {\it Spitzer} observations (0.6 days) is short in comparison with the typical Blazhko periods. Therefore, we retained the Blazhko variables in our analysis.

In order to clean our sample from blended sources and other contaminants we analysed the position of the RRLs  on the CMDs shown in Fig.~\ref{fig:cmd} %(right panels of Fig.~ \ref{fig:cmd}) 
and the light curves (Fig.~\ref{fig:lc1}) in combination with a visual inspection of the {\it Spitzer} images (Fig~\ref{fig:map2}). %The position of the RRLs on the CMD is shown in right panels of Fig.~\ref{fig:cmd}. 
Fig.~\ref{fig:cmd} demonstrates  that most of the  variables conform to the 
%RRab (red filled circles), RRc (blue triangles) and RRd (green squares) stars, with the 
RRc stars (blue triangles) being bluer than the RRab variables (red filled circles), and the RRd stars (green squares) being  located between RRc and RRab stars. However, for five objects, namely,  V01, V08, V19, V23 and V24 the position on the CMD is unusual. V08 is  far too bright  and red in colour than the bulk of the RRab stars. The star light curves are noisy, especially in the 4.5 $\mm$ band (Fig.~\ref{fig:lc1}). Inspection of the {\it Spitzer} images showed that  V08 is among an unresolved group of sources.  V01 is slightly separated and fainter than other RRLs in the CMD.  Moreover, V01 has a  significantly redder colour than it would be expected for its period (P=0.51~days). The source is  located on  the upper edge of the 3.6 and 4.5 $\mm$ images and does not have a characteristic stellar shape on the images, hence, the PSF photometry is likely not reliable. Star V23 is potentially a Blazhko variable and is slightly fainter than other RRLs, but the 3.6 and 4.5 $\mm$ images and the star light curves do not have particular issues. The faint magnitude of this star may be consistent with its period (0.46863 days) that is the shortest among all the  RRab stars in our sample.  It is worth noting that V23 has also the faintest $V$ \citep{Kuehn2013}  magnitudes among all RRLs in  Reticulum, which in combination with the results of this study may prove that V23 is indeed the intrinsically faintest star in the sample. While the faint MIR magnitudes can be explained by the shortest period of this star among all the RRab stars in the sample, the reason of the faint visual magnitude is unclear. A high metallicity of V23 could potentially account for its faint $V$ magnitude, however, it would contradict the general picture of Reticulum being monometallic cluster. Spectroscopical analysis of RRLs in Reticulum can shed light on this issue in the future. An alternative possibility is that V23 is a foreground RRL belonging to the field of the LMC. Star V19 has a too blue colour for an RRab variable. Moreover, it has a rather noisy light curve especially in the 4.5 $\mm$ band. Visual inspection of the images showed that V19 has a companion with 3.6 $\mm$ apparent magnitude  18~mag  at a distance of  0.82 arcsec. The source is also located close to the edge of the 4.5 $\mm$  image. Both these issues may have affected the photometry. V24 appears too red in colour for a RRd star and has an extremely noisy light curve.  The star is located at a distance of 3.79 arcsec from a 15~mag bright source, with which it is blended in both passbands. Finally, V28 and V32 are not separated from the other RRLs on the CMD, but have noisy light curves.  V28 has a close companion (at a distance of 1.63 arcsec) of magnitude $\sim 19$~mag, while V32 is blended by a companion with  magnitude 18~mag located at a distance of 1.87 arcsec.  Two further stars (V04 and V12)  have noisy light curves but their images and the position on the CMD do not indicate any clear issue. V04 is a double mode  pulsator and V12 is the shortest period (and hence faint) RRL in our sample. This may explain the noisy light curves. To summarise, the following six stars were dropped:  V01, V08, V19, V24, V28, V32, while V04 and V12 were retained to avoid biasing our derivation of the $PL$ relations by brighter/longer period stars. Therefore 
 in the following analysis we used a subsample of 24 RRLs consisting of  17 RRab, 3 RRc and 4 RRd stars.

\section{MIR $PL$ relations}\label{sec:pl}
\subsection{Slope of the PL relations.\label{sec:slope}}

\begin{figure}
    \includegraphics[trim=20 150 0 100,width=\linewidth]{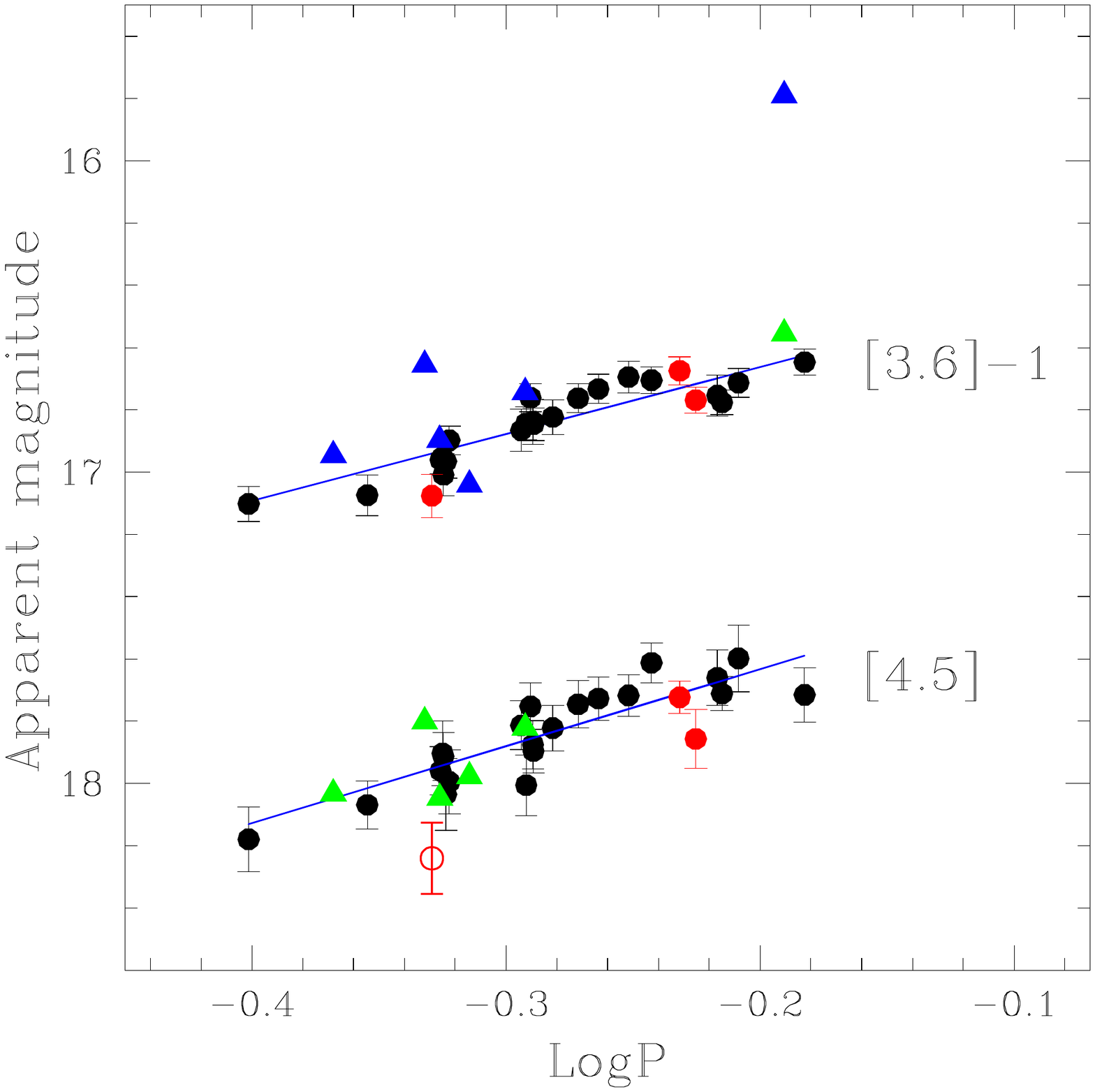}
  \caption{$PL$ relations in the 3.6 and 4.5 $\mm$ passbands defined by the Reticulum RRLs.  Filled circles represent RRLs used in the fit, with the Blazhko variables highlighted in red. Blue and green triangles show six RRLs that were discarded from the fit of respectively 3.6 and 4.5 $\mm$ $PL$ relations as a results of the analysis of the light curves, position on the CMD and {\it Spitzer} images (Section~\ref{subsec:rr}). Star V23 was also discarded from the fit of the 4.5 $\mm$ $PL$ relation by 3-sigma clipping procedure and is marked with a red empty circle (see text for details).}
  \label{fig:pl}
\end{figure}
%
%
%\begin{figure}
% % \includegraphics[trim=20 150 0 100,width=\linewidth]{PC_new.pdf}
%    \includegraphics[trim=20 150 0 100,width=\linewidth]{Fig8.pdf}
%  \caption{$PC$ relations defined by the Reticulum RRLs.  Filled circles represent 24 RRLs used in the fit, while red triangles mark six RRLs discarded from the analysis. Blue filled and dashed lines represent the $PC$ relation (Eq.~\ref{eq:col}) and one-sigma bounds, respectively. See text for the details.}
%  \label{fig:pc}
%\end{figure}

\begin{figure}
\includegraphics[trim=30 160 20 100, width=\linewidth]{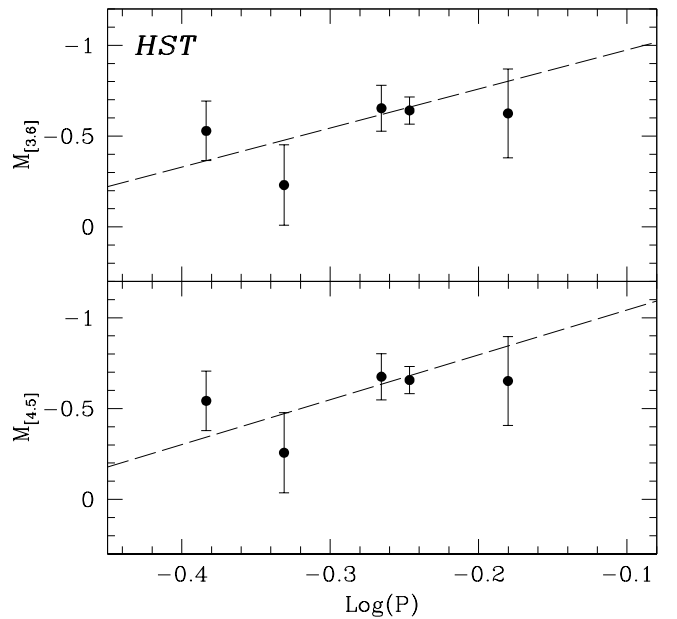}
  \caption{$PL$ relations in the 3.6 (upper panel) and 4.5 $\mm$ (lower panel) passbands of five Galactic RRLs with the zero points calibrated using the {\it HST} parallaxes of \citet{Benedict2011}. The logarithm of the period  of RZ Cep was ``fundamentalized'' (Eq.~\ref{eq:fund}). Slopes are from the present study (Eqs.~\ref{eq:M3_6}-\ref{eq:M4_5}).  The fits were performed in parallax space. See text for details.}
  \label{fig:pl_hst}
\end{figure}

\begin{figure}
    \includegraphics[trim=30 160 20 100, width=\linewidth]{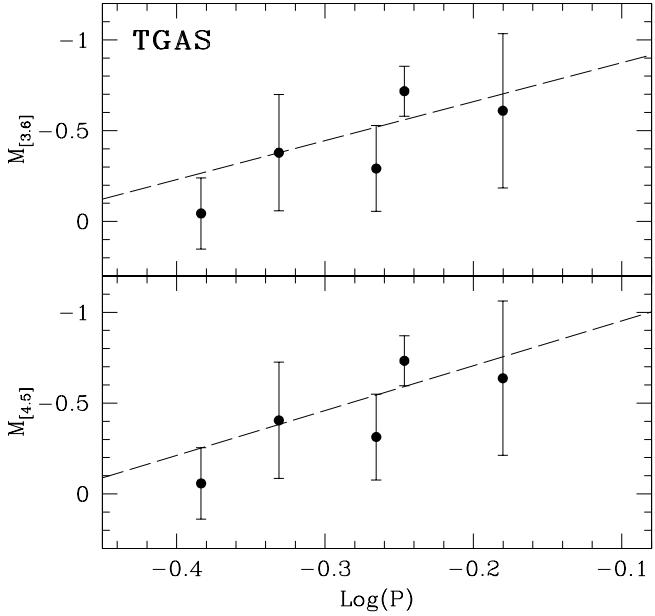}
  \caption{Same as in Fig.~\ref{fig:pl_hst}, but with the $PL$ %relations in the 3.6 (upper panel) and 4.5 $\mm$ (lower panel) passbands of five Galactic RRLs from  \citet{Benedict2011} with the 
  zero points calibrated using TGAS parallaxes for the  five Galactic RRLs. 
%  . The logarithm of period  of RZ Cep was ``fundamentalized'' by adding 0.127. Slopes are from this study of RRLs in Reticulum. Fit was performed in the parallax space. 
See Section~\ref{sec:zp} for details.}
  \label{fig:pl_tgas}
\end{figure}

\begin{figure}
    \includegraphics[trim=30 160 20 100, width=\linewidth]{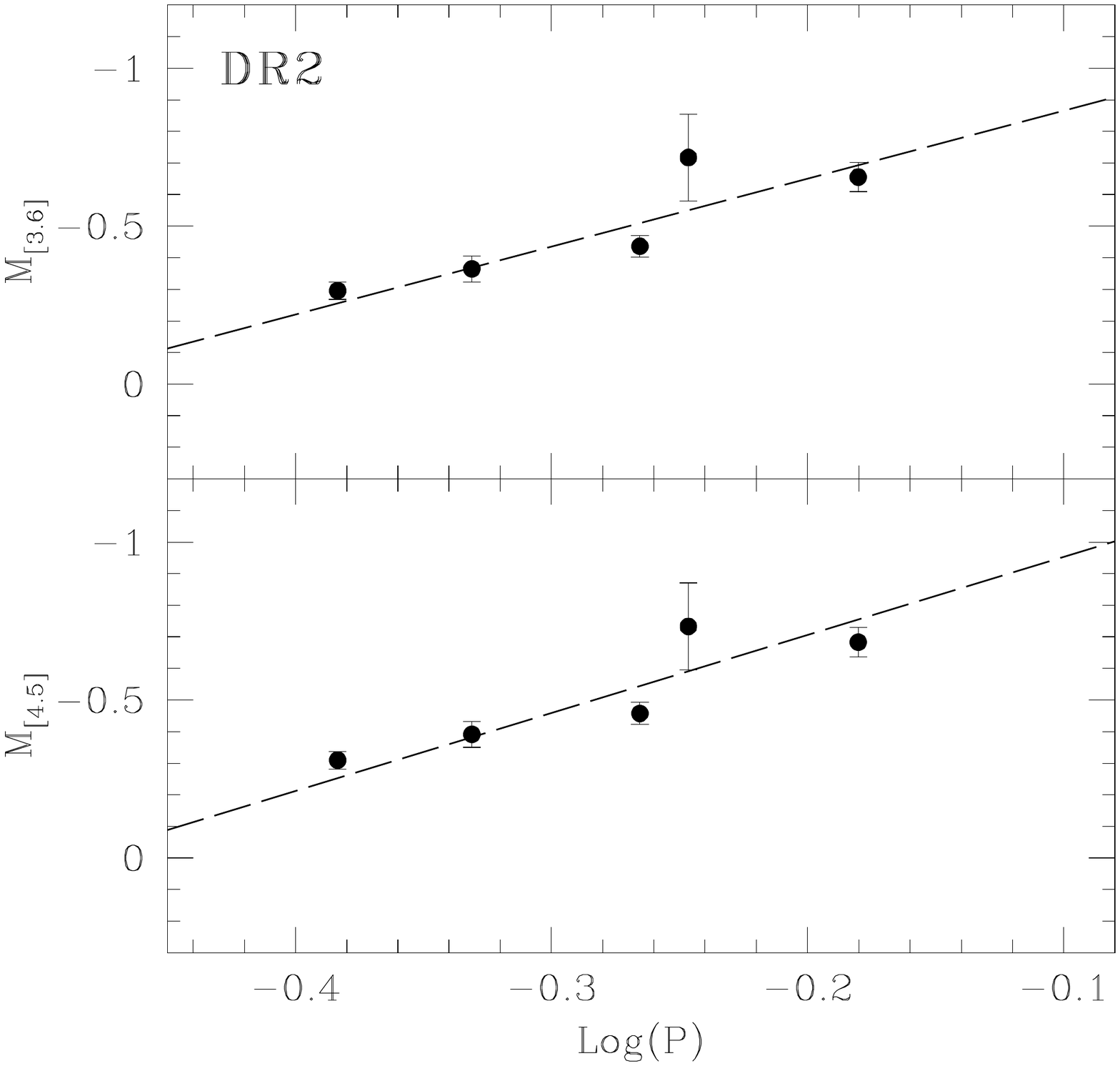}
  \caption{Same as in Fig.~\ref{fig:pl_hst}, but with the $PL$   zero points calibrated using DR2 parallaxes for the four Galactic RRLs and TGAS parallax for RR Lyr itself. See Section~\ref{sec:zp} for details.}
  \label{fig:pl_dr2}
\end{figure}

A number of studies indicate a decreasing dependence of the RRL luminosity on metallicity when moving to longer wavelengths. A rather small metallicity dependence of the $K$-band luminosity has been found by empirical studies  (see table~3 in \citealt{Muraveva2015} for a summary), while theoretical studies suggest the dependence of the $K$-band  luminosity on metallicity to be non-negligible (\citealt{Bono2001}, \citealt{Catelan2004}, \citealt{Marconi2015}), which recently was also confirmed by \citet{Braga2018} based on the analysis  of RRLs in the GC $\omega$~Cen. The MIR $PL$ relations were derived neglecting the dependence of the luminosity on metallicity  in a number of studies  (e.g. \citealt{Madore2013}; \citealt{Klein2014}; \citealt{Neeley2015},  \citealt{Clementini2017}). At the same time, some theoretical \citep{Neeley2017} and empirical (\citealt{Dambis2014}, \citealt{Sesar2017}, \citealt{Muraveva2018})  studies suggest a non-negligible dependence of the MIR luminosity on metallicity (metallicity slope $\ge 0.1$ mag/dex, see Table~\ref{tab:lit}). 
%Thus, a non-negligible dependence of the luminosity in the infrared bands on metallicity, that had been claimed by theoretical studies, also got confirmation by the empirical studies in the last years. 
The RRLs analysed  in this paper share the same metal abundance, hence, they cannot be used to estimate the effect of metallicity on the MIR $PL$ relations, however, we urge the reader to keep in mind a possible dependence of the MIR luminosity on metallicity.

%However, according to the monotonic sensitivity decrease with increasing the  wavelength we can safely assume that the metallicity dependence of the 3.6 and 4.5 $\mm$  $PL$ relations is negligible as it is also confirmed by \citet{Madore2013} and \citet{Neeley2015}. 
 
 Before we fit the MIR $PL$ relations, we corrected the apparent  3.6 and 4.5 $\mm$ magnitude of the RRLs in our sample for interstellar extinction. Reddening  $E(B-V)$ values towards the Reticulum cluster reported in the literature are generally low but controversial, and span the range from  0.016 mag \citep{Schlegel1998} to 0.06~mag \citep{Marconi2002}.
The range of reddening values for Reticulum reported in the literature  implies an uncertainty in the $V$ magnitude, and therefore its distance estimation,  of $\sim$ 0.14 mag. The situation improves significantly when moving to the MIR passbands, as the full range in reddening values reported in the literature causes  uncertainties of only 0.009 and 0.007 mag in the 3.6 and  4.5 $\mm$ bands, respectively,  which are negligible in comparison with other uncertainties. In this paper we adopt the reddening value $E(B-V)=0.03\pm0.02$ mag from \citet{Walker1992}, who estimated the reddening from the colours of the RRab stars at minimum light and Sturch's method (\citealt{Sturch1966}, \citealt{Walker1990}). This value of the reddening is close to the mean of the reddening estimates in the literature. The following relations obtained by \citet{Monson2012} based on the reddening laws from  \citet{Cardelli1989} and \citet{Indebetouw2005}, were used to estimate the interstellar extinctions: 
\begin{equation}
A_{[3.6]}=0.203*E(B-V)\label{eq:red3_6}
\end{equation}
\begin{equation}
A_{[4.5]}=0.156*E(B-V)\label{eq:red4_5}
\end{equation}
and derive the extinction values  of 0.006 and 0.005~mag in the 3.6 and  4.5 $\mm$ bands, respectively.
The  reddening-corrected 3.6 and 4.5 $\mm$ magnitudes were then used to fit the MIR $PL$ relations.

We used the weighted least squares method  and a 3-sigma clipping procedure to fit the $PL$ relations in the 3.6 and 4.5 $\mm$ passbands defined by the RRLs in our sample.  Following \citet{DallOra2004} the RRd variables were treated as RRc stars. Periods of first-overtone and double-mode RRLs were then ``fundamentalized'' according to the relation:
\begin{equation}
\log(P) = \log(P_{fo}) + 0.127\label{eq:fund},
\end{equation}
where $P_{fo}$ is the first-overtone period. The following $PL$ relations were derived:
\begin{equation}
[3.6] = (-2.15 \pm 0.23)\times \log(P) + (17.23 \pm 0.06)\label{eq:M3_6}
\end{equation}
with rms=0.06 mag, 
\begin{equation}
[4.5]=(-2.47 \pm 0.32)\times \log(P) + (17.14 \pm 0.09)\label{eq:M4_5}
\end{equation}
with rms=0.07 mag,\\
where [3.6]  and [4.5] are the absorption-corrected apparent magnitudes.
The best fit relations are shown in Fig.~\ref{fig:pl}, plotting with filled circles the RRLs used to fit the $PL$ relations, highlighting in red the Blazhko variables and marking with green and blue triangles  six stars that were discarded when fitting the 4.5 and 3.6 $\mm$ $PL$ relations, respectively, according to the analysis described in Section~\ref{subsec:rr}.  We also note that star V23 was automatically rejected by the 3-sigma clipping procedure when fitting the $PL$ relation in the 4.5 $\mm$ passband, which therefore is based on only 23 RRLs. We have marked this star with an empty circle in Fig.~\ref{fig:pl}.

%The Blazhko effect does not affect the determination of the average magnitudes of the stars in the sample since 
%The Blazhko variables are well fit by Eqs.~\ref{eq:M3_6} and \ref{eq:M4_5}  confirming that  the Blazhko effect does not significantly affect the mean magnitudes we measured for these stars.

%\vspace{-4mm}
\subsection{Zero point calibration\label{sec:zp}}

\begin{table*}
\begin{minipage}{16cm}
\tiny
%\begin{center}
 \caption{Comparison of MIR period-luminosity-(metallicity) relations in the form $\alpha\times \log(P) + \beta\times {\rm [Fe/H]} + ZP$}.\label{tab:lit}

\begin{tabular}{|lccccccc}
%\begin{tabular}{|||c|c|c|c@{}|}

\hline\hline
Reference & Band & $\alpha$ & $\beta$ & $ZP$   & $ZP$  & $ZP$  & $ZP$ from \\
{} & {} & {} & {} & from {\it HST} $\varpi$  & from TGAS  $\varpi$ & from {\it Gaia} DR2 $\varpi$ &other sources\\
%{} & {} & {} & {} &  \citet{Benedict2011} & other sources\\
%\hline
%~~~~~~~~~~~~~~~~~~~~~~~~~~~~~~~~~~~~~~~~~~~~~~~~~~~~~~~~~~~~~~~~[3.6]/W1\\
\hline 
\multicolumn{8}{|c|}{$PL$ relations ($W1$,  [3.6])}\\
\hline
This paper & 3.6 $\mm$ & $-2.150\pm0.230$ & $-$ & $-1.190\pm0.050$& $-1.090\pm0.090$ &$-1.080\pm0.03$& $-$ \\
\hline
\citet{Madore2013} & $W1$  & $-2.440\pm0.950$ & $-$ & $-1.260\pm0.250$ & $-$ & $-$ & $-$\\
\citet{Klein2014} RRab &$W1$ &$-2.380\pm0.200$& $-$& $-$ & $-$ &  $-$ & $-1.113\pm0.013^{a}$\\
\citet{Klein2014} RRc  & $W1$  &$-1.640\pm0.620$&  $-$& $-$ & $-$ &  $-$  & $-1.042\pm0.031^{a}$ \\
\citet{Neeley2015} & 3.6 $\mm$ & $-2.332\pm0.106$ & $-$& $-1.176\pm0.080$ & $-$ & $-$& $-1.054\pm0.020^{b}$ \\
\citet{Clementini2017} & $W1$ & $-2.440$ & $-$ & $-$ & $-1.210\pm0.040$ & $-$ & $-$ \\
\citet{Neeley2017} & 3.6 $\mm$ &  $-2.300\pm0.110$ &  $-$ &  $-1.112\pm0.089$ & $-$ & $-$ & $-$ \\
\hline
\multicolumn{8}{|c|}{$PLZ$ relations ($W1$,  [3.6])}\\
\hline
\citet{Dambis2014} &$W1$ &$-2.381\pm0.097$& $0.096\pm0.021$ &$-1.150\pm0.077$ & $-$ &$-$ & $-0.829\pm0.093^{c}$ \\
\citet{Neeley2017} &$W1$ & $-2.247\pm0.018$ &  $0.180\pm0.003$ & $-$ &$-$&  $-$ & $-0.790\pm0.007^{d}$ \\
\citet{Neeley2017} & 3.6 $\mm$ &  $-2.251\pm0.018$ &  $0.180\pm0.003$ & $-$ &$-$ & $-$ & $-0.793\pm0.007^{d}$ \\
 \citet{Sesar2017} & $W1$ & $-2.470^{+0.74}_{-0.73}$ & $0.150^{+0.09}_{-0.08}$& $-$ & $-0.890^{+0.12}_{-0.10}$$^{e}$ & $-$ & $-$\\
 \citet{Muraveva2018} & $W1$ & $-2.450^{+0.88}_{-0.82}$& $0.160^{+0.10}_{-0.10}$ & $-$ & $-$ & $-0.910^{+0.36}_{-0.34} $ & $-$ \\
\hline
%~~~~~~~~~~~~~~~~~~~~~~~~~~~~~~~~~~~~~~~~~~~~~~~~~~~~~~~~~~~~~~~~[4.5]/W2\\
\multicolumn{8}{|c|}{$PL$ relations ($W2$,  [4.5])}\\
\hline
This paper & 4.5 $\mm$& $-2.470\pm0.320$ & $-$ & $-1.290\pm0.050$& $-1.200\pm0.080$ & $-1.200\pm0.030$& $-$  \\
\hline
\citet{Madore2013} & $W2$ & $-2.550\pm0.890$ & $-$ & $-1.290\pm0.230$ & $-$ & $-$ & $-$ \\
\citet{Klein2014} RRab & $W2$  & $-2.390\pm0.200$ & $-$ & $-$ & $-$ & $-$ & $-1.110\pm0.013^{a}$ \\
\citet{Klein2014} RRc  & $W2$  & $-1.700\pm0.620$ & $-$ & $-$ & $-$ & $-$ & $-1.057\pm0.031^{a}$\\
\citet{Neeley2015} & 4.5 $\mm$ & $-2.336\pm0.105$ & $-$ & $-1.199\pm0.080$& $-$ & $-$ & $-1.091\pm0.020^{b}$\\
\citet{Neeley2017} & 4.5 $\mm$ &  $-2.340\pm0.100$ &  $-$ &  $-1.139\pm0.089$ & $-$  & $-$ & $-$ \\
\hline
\multicolumn{8}{|c|}{$PLZ$ relations ($W2$, [4.5])}\\
\hline
\citet{Dambis2014} &$W2$  &$-2.269\pm0.127$ & $0.108\pm0.021$ & $-1.105\pm0.077$ & $-$ & $-$ & $-0.776\pm0.093^{c}$\\
\citet{Neeley2017} & $W2$  &  $-2.237\pm0.018$ &  $0.185\pm0.003$ &  $-$ & $-$ & $-$ & $-0.784\pm0.007^{d}$ \\
\citet{Neeley2017} & 4.5 $\mm$ &  $-2.239\pm0.018$ &  $0.185\pm0.003$ &  $-$ & $-$ & $-$ & $-0.785\pm0.007^{d}$ \\
 \citet{Sesar2017} & $W2$  & $-2.400^{+0.84}_{-0.82}$ & $0.170^{+0.10}_{-0.09}$& $-$ & $-0.947^{+0.11}_{-0.10}$$^{e}$ & $-$ & $-$\\
\hline

\end{tabular}
\medskip
%\end{center}

$^{a}$ Zero point derived from  the $M_{V}-{\rm [Fe/H]}$ relation of RRLs in combination with the $\it HST$ parallaxes of RRLs in \citet{Benedict2011}.\\
$^{b}$ Zero point calibrated by adopting the distance modulus of the GC M4.\\
$^{c}$ Zero point determined from the statistical-parallax analysis.\\
$^{d}$ Theoretically determined zero point.\\
$^{e}$  Zero point is from table~1 and eq.~4 in \citet{Sesar2017}, assuming $P_{ref}=0.52854$ days and ${\rm [Fe/H]}_{ref}=-1.4$ dex.

\end{minipage}
\end{table*}

To transform to  absolute magnitudes the apparent magnitudes in Eqs.~\ref{eq:M3_6} and \ref{eq:M4_5} we must calibrate the zero points of the $PL$ relations. 
%To use RRLsas distance indicators one needs to calibrate the zero points of the MIR $PL$ relations. 
Trigonometric parallaxes are the only direct method to  estimate distances and infer absolute magnitudes.  Until recently accurate parallax measurements  were available for only a handful of RRLs.  The ESA  mission {\it Hipparcos} (\citealt{vanLeeuwen2007}  and references therein) measured the  parallax for more than 100 RRLs in the solar neighbourhood, but with errors larger than 30\%,  except for  RR Lyr itself, the bright prototype of the RRL class,  for which  the error is $\sim18$\%. \citet{Benedict2011} measured accurate  trigonometric parallaxes  of five Galactic RRLs (among which is RR Lyr) using the FGS on board  the {\it HST}. However, we note that the {\it Hipparcos} and $HST$ parallax of  RR Lyr are only  in marginal agreement.
Therefore, a direct calibration of the RRL distance relations has so far been hampered not only  by a lack of good quality parallax data,  but also by the few independent measurements presently available providing inconsistent results. 
% itself is in contrast with the star parallax measured by the {\it HST}. Thus, studies into direct calibration of RR Lyrae distance relations were not only hampered by an extreme lack of good quality %parallax data, both in precision and quantity, but the few sources of parallax data available provided inconsistent results. 
{\it Gaia}, the ESA cornerstone  mission  launched on  2013  December 19, is measuring trigonometric parallaxes (along with positions, proper motions, photometry and physical parameters) for over one billion stars in the MW and beyond. {\it Gaia} will revolutionise the field by providing end-of-mission parallax measurements 
with about 10 $\mu as$ uncertainty for RRLs brighter than $V \sim$ 12-13 mag.  

A first  anticipation of {\it Gaia} promise in this field came on  2016 September 14 with the publication of the {\it Gaia} DR1 (\citealt{Prusti2016}, \citealt{Brown2016}). The DR1 catalogue contains parallaxes for about 2 million stars  in common between {\it Gaia} and the {\it Hipparcos}  and Tycho-2 catalogues, computed as part of the TGAS \citep{Lindegren2016}. The TGAS sample includes parallaxes for 364 MW RRLs, of which a fraction was used by \citet{Clementini2017} to calibrate the RRL  visual $M_{V}-{\rm [Fe/H]}$ relation and the infrared $PL$ and $PLZ$ relations. 
Even though TGAS parallaxes for RRLs show an extraordinary improvement  with respect to  {\it Hipparcos}  (see e.g. figure~23 in \citealt{Clementini2017}), their uncertainties are still  large, spanning from 0.209 to 0.967~mas in range. The situation improved significantly on 2018 April 25, when the {\it Gaia} trigonometric parallaxes for a much larger sample of RRLs using only {\it Gaia} measurements were published in {\it Gaia} DR2. The {\it Gaia} DR2 catalogue comprises positions and multi-band photometry for $\sim1.7$ billion sources alongside the parallaxes and proper motions for $\sim1.3$ billion sources  \citep{Brown2018}. Furthermore, {\it Gaia} DR2 published a catalogue of more than $\sim$500,000 variables of different types \citep{Holl2018}, including 140,784 RRLs  \citep{Clementini2018}. The dramatic improvement brought by {\it Gaia} DR2 for RRLs, compared to its precursors {\it Hipparcos} and {\it Gaia} DR1, is demonstrated in figure 6 of \citet{Muraveva2018}.

%An impressive improvement of quality and accuracy of {\it Gaia} DR2 parallaxes of RRLs in comparison with {\it Hipparcos} and TGAS data is shown, for instance, in figure~6 of \citet{Muraveva2018}.}

We retrieved {\it Gaia} DR2 parallaxes for the 30 RRLs in Reticulum from the {\it Gaia}  Archive website\footnote{\url{http://archives.esac.esa.int/gaia}}. The resulting values are shown in Table~\ref{tab:gen}. Unfortunately, owing to the large distances and faint magnitudes of the RRLs in Reticulum (mean {\it Gaia} $G$-band magnitude $<G> = 19.02$~mag), 50\% of the Reticulum RRLs have negative parallax values, while the remaining 15 RRLs have large relative parallax uncertainties ($<\sigma_{\varpi} / \varpi >= 2.7$). 
Consequently, the Reticulum sample is unsuitable for calibrating the zero points of the RRL MIR $PL$ relations. Instead we adopt the slope of the MIR $PL$ relations derived for the RRLs in Reticulum (Eqs.~\ref{eq:M3_6} and \ref{eq:M4_5}), and calibrate the zero points using the parallaxes of RRLs in the MW. We selected   the  {\it HST} parallaxes for the five RRLs in  \citet{Benedict2011} and the  TGAS and {\it Gaia} DR2  parallaxes for the same stars, and then compared  the resulting $PL$ relations with the literature relations (Section~\ref{subsection:comparison}).

% we adopt the slope of the MIR $PL$ relations derived for the RRLs in Reticulum (Eqs.~\ref{eq:M3_6} and \ref{eq:M4_5}), and the zero points calibrated using the parallaxes of RRLs in the MW. For the sake of comparison we selected {\bf  the  {\it HST} parallaxes for the five RRLs in  \citet{Benedict2011} and the  TGAS and {\it Gaia} DR2  parallaxes for the same stars,} and then compared  the resulting $PL$ relations with the literature relations (Section~\ref{subsection:comparison}).} 
%As a consequence, this sample can not be used to calibrate the zero point of the RRL MIR $PL$ relations. In this study we adopt the slope of the MIR $PL$ relations derived for the RRLs in Reticulum (Eqs.~\ref{eq:M3_6} and \ref{eq:M4_5}), and the zero points calibrated using the parallaxes of RRLs in the MW. For the sake of comparison we selected {\bf  the  {\it HST} parallaxes for the five RRLs in  \citet{Benedict2011} and the  TGAS and {\it Gaia} DR2  parallaxes for the same stars,} and then compared  the resulting $PL$ relations with the literature relations (Section~\ref{subsection:comparison}).}  

%In our study we decided to adopt the slopes of the $PL$ relations derived for 24 RRLs in Reticulum (Section~\ref{sec:pl}), 
Apparent magnitudes in the {\it Spitzer} passbands for the five MW RRLs with {\it HST} parallaxes published by  \citet{Benedict2011}, namely,
RZ Cep, XZ Cyg, SU Dra, RR Lyr and UV Oct, are provided in 
%(RZ Cep, XZ Cyg, SU Dra, RR Lyr and UV Oct) in {\it Spitzer} passbands from \citet{Neeley2015} in combination with the {\it HST} and TGAS trigonometric parallaxes of these stars in order to derive our own  $PL$ relations. Apparent magnitudes of the five RRLs are provided in 
table~5 of \citet{Neeley2015}, {\it HST} parallaxes are taken from table~8 of \citet{Benedict2011} and the TGAS and   {\it Gaia} DR2 parallaxes are retrieved from {\it Gaia}  Archive website but  they are  also listed in table~1 of \citet{Clementini2017}, for TGAS, and table~2 of \citet{Muraveva2018}, for the DR2 parallaxes.
Periods of these five stars range from 0.3086 to 0.6604 days and metallicities span the range from [Fe/H]=$-$1.80 to $-$1.41 dex on the \citet{Zinn1984} metallicity scale. The metallicity of the RRLs  in Reticulum is [Fe/H]=$-$1.66 dex on the \citet{Zinn1984} metallicity scale \citep{Mackey2004}, which is well within the range spanned  by \citet{Benedict2011}'s variables.  
The logarithm of the RRc star RZ Cep was ``fundamentalized'' (Eq.~\ref{eq:fund}).  

 The DR2 parallax for  RR Lyr itself  has a large negative value ($-2.61\pm0.61$~mas)  and is clearly wrong (\citealt{Arenou2018}, \citealt{Brown2018}). Hence, for this star we adopt the TGAS parallax instead of the DR2 measure.
%\citet{Benedict2011} provides two different parallax values for RZ Cep,  namely, 2.12 and 2.54~mas. Although \citet{Benedict2011} preferred value  is 2.12~mas, \citet{Neeley2015} adopted 2.54~mas  for RZ Cep parallax. \citet{Clementini2017} show in their figure~11 that the  2.54~mas parallax value is in much better agreement with the TGAS parallax of RZ Cep. Thus, in our study we also prefer to use $\varpi$= 2.54~mas for RZ~Cep. 

As widely discussed in \citet{Clementini2017},  the direct transformation of parallaxes  into absolute magnitudes  using the relation:

\begin{equation}\label{eq:abs_mag}
M = m_0 + 5\log \varpi -10, 
\end{equation}
where M is the star absolute magnitude, $m_0$ is the dereddened apparent  magnitude and $\varpi$ is the parallax in mas, has significant disadvantages.  While errors in parallaxes are approximately Gaussian and symmetrical, the logarithm in Eq.~\ref{eq:abs_mag} causes the errors in absolute magnitudes to become asymmetrical leading to biased results (\citealt{Clementini2017}; \citealt{Luri2018}). Furthermore, using this method means that negative parallaxes cannot be transformed into absolute magnitudes which causes a truncation bias in the data. In order to circumvent both issues it is advisable to operate directly in the parallax space, using, for instance, the reduced parallax  \citep{Feast1997} or Astrometry-Based Luminosity (ABL, \citealt{Arenou1999}) method that is defined by the relation:
\begin{equation}\label{eq:abl}
ABL = 10^{0.2M} = 10^{0.2(\alpha \log P + ZP)} =  \varpi10^{0.2m_0-2},
\end{equation}
where M is the stars absolute magnitude, P is the period, $m_0$ is the dereddened apparent  magnitude, $\varpi$ is the parallax in mas, $\alpha$ is the slope of the $PL$ relation, for which we adopt the values we have derived from the RRLs in Reticulum (Eqs.~\ref{eq:M3_6} and \ref{eq:M4_5} in Section~\ref{sec:slope}), and $ZP$ is  the $PL$ zero point that we aim to determine.
%we aim to calibrate using the {\it HST} parallaxes and, as an alternative,  the TGAS and \textbf{DR2} parallaxes of the five MW RRL calibrators. 
Using the ABLs instead of the absolute magnitudes for our RRL calibrators allows us to  maintain symmetrical errors.\footnote{More details on the use of the ABL method to  fit the $PL$ relation can be found in \citet{Clementini2017}.} 
We also note that there is no need to apply any Lutz - Kelker correction \citep{Lutz1973} in our study, since individual measurements of objects that were  chosen independently of their parallax values or relative errors in parallaxes are not affected by the Lutz - Kelker bias (\citealt{Arenou1999}, \citealt{Feast2002}), and indeed the five RRLs in \citet{Benedict2011} were not selected based on their observed parallaxes.

  A number of independent studies regarding a possible  zero point offset of {\it Gaia} DR2 parallaxes appeared recently in literature  (e.g. \citealt{Arenou2018}; \citealt{Riess2018}; \citealt{Zinn2018}; \citealt{Stassun2018}; \citealt{Muraveva2018}). All these studies indicate that  the {\it Gaia} DR2 parallaxes are systematically smaller with a mean estimated offset of $\Delta \varpi = -0.03~mas$ \citep{Arenou2018}.  \citet{Arenou2018} do not recommend to correct for this possible parallax offset individual star parallaxes. Hence, we did not apply this correction to the five MW RRLs in our study. 
%It is also worth noting that owing to the non-linear relation between parallax and absolute magnitude a given parallax offset does not affect all absolute magnitudes equally and is much more pronounced for the farther stars \citep{Muraveva2018}. However, since  the five  RRLs from \citet{Benedict2011} are relatively close to us we don't expect a strong effect of a possible DR2 parallax offset on the derived absolute magnitudes.}

The resulting 3.6 and 4.5 $\mm$ $PL$ fits obtained by using the {\it HST}, TGAS and  DR2 parallaxes of the five MW RRL calibrators are shown in Figs.~\ref{fig:pl_hst}, \ref{fig:pl_tgas} and \ref{fig:pl_dr2}, respectively.  The fits were performed in parallax  space, with parallaxes later transformed to absolute magnitudes in order to visualise the data on the  $PL$ plane. The slopes and zero points of the fits are listed in Table~\ref{tab:lit}.

\subsection{Comparison with the literature}\label{subsection:comparison}
In recent years, several authors have studied the RRL  MIR  $PL$ relations.
These literature relations  were calibrated either using the trigonometric parallaxes or by indirect methods to determine the zero point. Here, we briefly summarise these previous studies with a specific focus on the adopted calibration procedures. In Table~\ref{tab:lit}  we present the literature slopes and zero points alongside the values derived in the present study.\\

\citet{Madore2013} used {\it Wide-field Infrared Survey Explorer (WISE)} observations in the $[W1]$ (3.4 $\mu$m), $[W2]$ (4.6 $\mu$m) and $[W3]$ (12 $\mu$m) passbands for  the four fundamental mode RRLs in \citet{Benedict2011} (they excluded the first-overtone star RZ Cep)  
to derive $PL$ relations in the {\it WISE} MIR passbands directly  calibrated on the  {\it HST} parallaxes of the  stars. 

Similarly, \citet{Dambis2014} derived metallicity-dependent MIR $PLZ$ relations  %in the {\it WISE} MIR passbands  
using {\it WISE}  $W1$ observations for 360 RRLs in 15 MW GCs and $W2$ observations  for  275 RRLs in 9 GCs. The authors calibrated the zero points in two different ways: based on a statistical-parallax analysis and using the {\it HST} parallaxes of the four RRLs in \citet{Madore2013}. The corresponding zero points differ by  more than  0.3~mag, with the statistical-parallax calibration providing fainter absolute magnitudes. %the {\it HST} absolute magnitudes being brighter. %The MIR $PL$ relations found by \citet{Dambis2014} are slightly metallicity dependent. 
$PL$ relations in the {\it WISE}  $W1$, $W2$ and $W3$ passbands were derived also by \citet{Klein2014}  %derived $PL$ relations 
for RRab and RRc stars separately, using 129  field MW RRLs whose distances were inferred from the ${M_{V}-{\rm [Fe/H]}}$ relation in \citet{Chaboyer1999} calibrated with \citet{Benedict2011}'s trigonometric parallaxes.

\citet{Neeley2015} derived $PL$ relations in the 3.6 and 4.5 $\mm$ passbands using 37 RRLs  observed with {\it Spitzer} IRAC in the GC M4. The zero point calibration was obtained combining the {\it HST} parallaxes with the {\it Spitzer} photometry of  the 5 RRLs  in  \citet{Benedict2011}  observed as part of the CRRP and, as an alternative,  by adopting the distance to M4 from \citet{Braga2015}. New theoretical $PLZ$ relations in the  {\it Spitzer} and {\it WISE} passbands were published by 
 \citet{Neeley2017} who also revised the empirical relations in  \citet{Neeley2015}  using  M4 data reduced with a new {\it Spitzer} pipeline.

 \citet{Clementini2017} calibrated their own $PL$ relation in the $W1$ passband using the TGAS parallaxes of 198 MW RRLs and adopting the slope from \citet{Madore2013}. \citet{Clementini2017} used three different approaches to infer the zero point of the $PL$ relations. In Table~\ref{tab:lit} we list the zero point derived using the ABL (\citealt{Arenou1999}), which is also the procedure adopted to calibrate  the $PL$ relations derived in the present paper. 
 
  \citet{Sesar2017} studied a sample of about 100 MW RRab stars and calibrated the  MIR $W1$ and $W2$ $PLZ$ relations based on TGAS parallaxes. Recently, \citet{Muraveva2018} derived new $W1$-band $PLZ$ relations using {\it Gaia} DR2 parallaxes for a sample of 401 MW RRL variables and for a reduced sample of 23 nearby RRLs whose metallicity was estimated from high-resolution spectroscopy. 
 %In that study we urged the readers that there is a severe correlation between the possible offset in  {\it Gaia} DR2 parallaxes and the metallicity slope of the $M_V - {\rm [Fe/H]}$, $PM_KZ$ and $PW1Z$ relations of RRLs. However, the effect of the parallax offset on the metallicity slope is significantly reduced for the close stars. 
 The MIR relation obtained by \citet{Muraveva2018} using the sample of 23 nearby RRLs after correcting for a {\it Gaia} DR2 parallax offset of 0.056~mas is shown in Table~\ref{tab:lit}. For ease of comparison  in Table~\ref{tab:lit} we grouped the $PL$ and $PLZ$ relations  in the $W1$ and  3.6 $\mm$ passbands, and those in the  $W2$ and  4.5 $\mm$ bands, since they are very similar.

The  slope of the $PL$ relation in the 3.6 $\mm$ band derived in this paper is shallower than  values published in previous studies but still consistent with them within the errors. The slope of the $PL$ relation in 4.5 $\mm$ passband is in  very good agreement, within the quoted uncertainties, with all previous studies.  The absolute magnitudes based on the TGAS and DR2 parallaxes are in very good agreement to each other and  generally  fainter than those based on the {\it HST} parallaxes.
 In Section~\ref{sec:dist} we discuss the distance to Reticulum  inferred from the new $PL$ relations and their  different  calibration procedures.
%We envisage re-deriving new zero points for Eqs.~\ref{eq:M3_6} and \ref{eq:M4_5} when improved parallax values for the five {\it HST} RRL calibrators  may become available with {\it Gaia} DR2.

\begin{table*}
\begin{minipage}{18cm}
%\begin{center}
 \caption{Comparison of the distance moduli for Reticulum derived in the present study and literature values.}\label{tab:dist}

\begin{tabular}{lcccl}
%\begin{tabular}{|||c|c|c|c@{}|}

\hline\hline
~~~~~~~~~~~~~~~Reference ~~~~~~~~~~~~~~~&  ~~~~~~~~~~~~~~Band ~~~~~~~~~~~~~~~& $E(B-V)$ & ~~~~~~~~~~~~~~~~~~~Distance modulus~~~~~~~~~~~~\\ %\hline
{}&{}&(mag) & (mag)\\
%~~~~~~~~~~~~~~~~~~~~~~~~~~~~~~~~~~~~~~~~~~~~~~~~~~~~~~~~~~~~~~~~[3.6]/W1\\
\hline 
\citet{Demers1976} & $V$ & 0.02  & 18.51 \\
\citet{Walker1992} & $V$ & 0.03  & 18.38\\
%\citet{Bono2003} & $V$ & & \\
\citet{Ripepi2004}  & $V$ & 0.02  & $18.39\pm0.12$  \\
\citet{Mackey2004}  & $V$ & 0.05 & $18.39\pm0.12$\\
\citet{Kuehn2013} & $V$ & 0.016 & $18.40\pm0.20$ \\
\citet{Jeon2014} & $V$ & 0.03& 18.40\\
\citet{Sollima2008} & $V$ & 0.03& $18.44\pm0.14$\\
\hline
\citet{Kuehn2013} & $I$ & 0.016  & $18.47\pm0.06$ \\
\hline
\citet{DallOra2004} & $K$ &0.03 & $18.52\pm0.05  \pm 0.117 $\\
\citet{Sollima2008} & $K$ & 0.03 & $18.48\pm0.11$\\
\hline 
This paper ({\it HST}) &  [3.6] & 0.03 & $18.43\pm0.06$ \\
This paper (TGAS) &  [3.6] & 0.03 & $18.33\pm0.06$ \\
This paper (DR2) &  [3.6] & 0.03 & $18.32\pm0.06$ \\
This paper ({\it HST}) &  [4.5] & 0.03 & $ 18.43\pm0.08$ \\
This paper (TGAS) &  [4.5] & 0.03 & $18.34\pm0.08$\\
This paper (DR2) &  [4.5] & 0.03 & $18.34\pm0.08$\\

\hline

\end{tabular}
\medskip
%\end{center}

\end{minipage}
\end{table*}

%\vspace{-6mm}
\section{Distance to Reticulum}\label{sec:dist}
Several independent estimates of the distance to  Reticulum are available in the literature. Table~\ref{tab:dist} summarises the distances derived in this study and other key works. We indicate the photometric passbands and the reddening values used in the various analyses.

\citet{Demers1976} derived a distance modulus for Reticulum of $\mu=18.51$ mag using  the mean apparent $V$ magnitudes of the HB stars. These authors also found that the mean $B$ magnitude of the RRLs in Reticulum is about 0.15 mag brighter than for RRLs in the  LMC field.  Several independent studies, such as  \citet{Freedman2001} based on classical  Cepheids, \citet{Clementini2003} using RRLs and \citet{Pietrzynski2013} from a sample of eight eclipsing binaries, have shown that the distance to the LMC is $\mu\sim18.50$ mag (see e.g \citealt{DeGrijs2014} for a compilation of literature values). Therefore,  the 0.15 mag difference in mean $B$ magnitude  found by \citet{Demers1976} implies $\mu \sim18.35$ mag for the distance modulus of Reticulum,  assuming that the reddening and $B$ absolute magnitude are the same in the two systems.

\citet{Walker1992}, \citet{Ripepi2004}, \citet{Mackey2004}, \citet{Kuehn2013}  derived the distance to Reticulum  from the $V$-band observations  of the 32 RRLs in the cluster. 
 \citet{Kuehn2013} also derived a distance modulus for Reticulum using their $I$-band photometry of the cluster RRL variables, individual metallicities estimated from the Fourier parameters of the light curves and adopting the $PLZ$ relation of \citet{Catelan2004}. 
%Individual metallicities of the RRLs were estimated from the Fourier analysis of the light curves ({\bf se lasci questa frase devi mettere anche la  referenza, dire che valore medio viene fuori dai parametri di Fourier e come si confronta con le metallicita' spettroscopica di Reticulum se e' mai stata stiamata.}) 
\citet{Jeon2014} derived the distance to Reticulum by fitting theoretical isochrones to the cluster $V$ versus ($B-V$) CMD. All distance moduli based on the $V$ photometry agree on a distance modulus value around $\mu=18.40$ mag, except  \citet{Demers1976}'s that is about 0.1 mag  larger.

%,  except the one by \citep{Demers1976}, cluster around the value of $\mu=18.40$ mag. 
\citet{DallOra2004} combined  the apparent $K$-band  magnitudes of 30 RRLs in Reticulum with the theoretical  $PLZ$ relation by \citet{Bono2003}  obtaining  the distance modulus $\mu=18.52\pm0.005(random)\pm0.117(systematic)$ mag which is about 0.1 mag  larger than that derived from the $V$ photometry. These authors made a detailed analysis of possible systematic uncertainties affecting  their analysis. Distances obtained from the NIR $PLZ$ relation were found to be systematically larger than distances derived from  Baade-Wesselink studies \citep{Fernley1994}, with the difference ranging from  $\delta \mu=0.10$ mag for more  metal-rich clusters to $\delta \mu=0.32$ mag for  more metal-poor clusters. Therefore, the distance modulus of $\mu=18.52$ mag derived from the $K$-band photometry  of Reticulum  could be an overestimate. Later,  \citet{Sollima2008} also found that the distance to Reticulum derived from the $K$-band photometry of \citet{DallOra2004} is 0.04 mag  larger than the distance inferred from the $V$-band photometry of \citet{Walker1992}. 

%To analyse the systematic uncertainties affecting the distance estimates \citet{DallOra2004} selected four Galactic GCs for which optical and NIR photometry are available. The authors %compared the distance moduli of the GCs derived by applying the NIR $PLZ$ relation \citep{Bono2003} with the distances based on the Baade-Wesselink calibration provided by 
%\citet{Fernley1994} and based on the first-overtone blue edge from \citet{Caputo2000}. \citet{DallOra2004} found that the distances obtained from the NIR $PLZ$ relation are systematically %larger than the distances derived from the Baade-Wesselink studies with difference ranging from  $\delta \mu=0.10$ mag for the more  metal-rich clusters to $\delta \mu=0.32$ mag for the more %metal-poor clusters, hence, the distance modulus of Reticulum derived from the $K$-band photometry $\mu=18.52$ could be overestimated. 

% \citet{Sollima2008} derived the absolute $V$ and $K$ magnitudes of RR Lyrae itself and used these values in combination with the $V$-band \citep{Walker1992} and $K$-band photometry \citep{DallOra2004} of RRLs in Reticulum to obtain the distance moduli of the cluster. The distance derived from the $K$-band photometry is 0.04 mag larger than the one derived from the $V$-band data. 
A comparison of the values  in  Table~\ref{tab:dist} indeed confirms that distances based on $K$- and $I$-band photometry are in general  larger than  distances derived from the optical photometry,  which could suggest some issues in  the  zero point calibrations in the two different passbands. 
 %The situation will improve when the {\it Gaia} parallaxes will become available and the absolute magnitudes of RRLs in all passbands will be derived with an unprecedented accuracy.  
 
%As discussed in Subsection~\ref{subsec:red} the value adopted for the reddening in Reticulum differs significantly from study to study hence  affecting the derived distance moduli particularly when %optical ($B$ and $V$) photometry is used. 

%To avoid the reddening issue we used our own MIR $PL$ relations in 3.6 and 4.5 $mm$ passbands calibrated using {\it HST} and TGAS parallaxes (Section~\ref{sec:zp})  

 In this study we obtained  distance moduli for Reticulum  by applying our $PL$  relations to estimate  individual distances to 
each of the  24 and 23 RRLs   used  in our analysis of the 3.6 $\mm$ and 4.5 $\mm$ $PL$ relations, respectively, then averaging the resulting values. Errors are the standard deviations from the average.
We obtained the following distance moduli for Reticulum:  $\mu_{[3.6]}=18.43\pm0.06$~mag  and $\mu_{[4.5]}= 18.43\pm0.08$~mag from the $PL$ relations calibrated  on the  {\it HST} parallaxes; $\mu_{[3.6]}=18.33\pm0.06$~mag and $\mu_{[4.5]}=18.34\pm0.08$~mag  using the TGAS parallaxes and $\mu_{[3.6]}=18.32\pm0.06$~mag  and $\mu_{[4.5]}=18.34\pm0.08$~mag using the DR2 parallaxes. These values are also reported in the bottom part of Table~\ref{tab:dist}. 
The distance moduli estimated using the $PL$ relations calibrated on the  {\it HST} parallaxes are in good agreement with the majority of the literature  estimates, especially with those based on the $V$ photometry. Distance moduli obtained  from  the $PL$ relations calibrated on the TGAS and DR2 parallaxes are generally smaller, but still consistent within errors with the majority of previous estimates. 

\citet{Pietrzynski2013}  estimated a distance to the LMC of $D = 49.97 \pm 0.19$ (stat) $\pm$ 1.11 (syst) kpc ($\mu_0$= 18.493 $\pm$ 0.008 (stat) $\pm$  0.047 (syst) mag) from eight long-period eclipsing binary systems which are all  located relatively close to the barycentre of the LMC. Taking the mean value of all six distance moduli derived in this study we conclude that Reticulum is located approximately 3 kpc closer to us than the barycentre of the LMC.

%\begin{figure}
%  \includegraphics[width=\linewidth]{Slopes.ps}
%  \caption{VMC coverage of the SMC, where $\alpha_0$ = 12.5 deg, $\delta_0$ = -73 deg. Blue boxes are VMC tiles, red boxes represent VMC tiles selected for the current study. Black dots are RRLs observed by the OGLE~III survey.}
%  \label{fig:slopes}
%\end{figure}

\section{Summary}\label{sec:sum}

We have analysed a sample of  30 RRLs in the LMC cluster Reticulum that we  observed in the 3.6 and 4.5 $\mu m$ passbands with IRAC  in the {\it Spitzer} Warm Cycle 9 as part of the CRRP. Observations consist of 12 epochs  over  a $\sim14$ hours time span,  allowing a good coverage of the RRL  light curves. We analysed the light curves with the GRATIS software to derive MIR characteristic parameters  (amplitudes and intensity-averaged mean magnitudes). We combined our MIR data with  optical photometry from \citet{Jeon2014}, and the resulting CMDs show a well  defined RGB, an HB  well populated by  the cluster RRLs and a few variables with magnitude/colour affected by crowding.
%, which confirms the quality of the photometry and calibration we performed.
Based on our analysis of the light curves, the position on the CMD and the appearance on the images we selected a sample of 24 RRLs  that we used to fit the MIR $PL$ relations. The slope of the $PL$ relation in 3.6 $\mm$ passband derived in this paper ($ -2.15 \pm 0.23$) is shallower than the values  published in previous studies but still consistent within the errors. The slope of the $PL$ relation in the 4.5 $\mm$ passband  ($ -2.47\pm 0.32$),  based on the 23 RRLs remained after the 3-sigma clipping procedure,  is in very good  agreement, within the errors, with the literature studies.

 To estimate the zero points of our MIR $PL$ relations we used five MW RRLs for which trigonometric parallaxes were measured with the {\it HST} FGS \citep{Benedict2011}, {\it Gaia} as part of TGAS  in {\it Gaia} DR1 and {\it Gaia} DR2, based only on {\it Gaia} measurements,  allowing us to make a direct comparison of the three calibrations. 
We used the ABL method to fit the MIR RRL $PL$ relations, as this operates directly in parallax space, avoiding non-symmetrical errors in absolute magnitudes caused by inversion of the parallaxes. We found that the absolute magnitudes based on the TGAS and  DR2 parallaxes are in good agreement with each other and are in general fainter than those based on the {\it HST} parallaxes. However, the sample of RRLs used in the analysis is too small to draw any conclusion about possible systematic offsets. 

 We obtained the following distance moduli for Reticulum:  $\mu_{[3.6]}=18.43\pm0.06$~mag  and $\mu_{[4.5]}= 18.43\pm0.08$~mag from the $PL$ relations calibrated  on the  {\it HST} parallaxes; $\mu_{[3.6]}=18.33\pm0.06$~mag and $\mu_{[4.5]}=18.34\pm0.08$~mag  using the TGAS parallaxes and $\mu_{[3.6]}=18.32\pm0.06$~mag  and $\mu_{[4.5]}=18.34\pm0.08$~mag using the DR2 parallaxes. These distance moduli  are in good agreement  with the literature values estimated using the $V$ band photometry, while the agreement with the distance values of Reticulum based on the $I$ and $K$ photometry is less pronounced.

Based on the mean of the distance moduli derived in this study we conclude that Reticulum is   located approximately 3 kpc closer to us than the barycentre of the LMC. 

%separated from the disk of the LMC and located $\sim3$ kpc closer to us. 

%We found that the distance moduli of Reticulum obtained from the $K$ and $I$ photometry are generally $\sim0.1$ mag larger than the distance moduli  based on the $V$-band and MIR observations. This difference could be related to the zero point calibrations of the $M_{V}-{\rm [Fe/H]}$ and $PL$ relations in different passbands. This issue will be resolved when the ESA cornerstone mission Gaia will provide the unprecedentedly accurate parallaxes for 100-150 Galactic RRLsthat will allow to calibrate the $M_{V}-{\rm [Fe/H]}$ and infrared  $PL$ relations with a great accuracy. These relations will be used to determine accurate distance to the Local group galaxies end improving the whole cosmic distance ladder. 

%\vspace{-5mm}
\section*{Acknowledgments}
 We thank the anonymous reviewer for a careful reading and valuable suggestions that have improved the paper.
This work is based on observations made with the Spitzer Space Telescope, which is operated by the Jet Propulsion Laboratory, California Institute of Technology under a contract with NASA. Support for this work was provided by NASA through an award issued by JPL/Caltech and 
makes use of data from the European Space Agency (ESA)
mission {\it Gaia} (\url{https://www.cosmos.esa.int/gaia}), processed by
the {\it Gaia} Data Processing and Analysis Consortium (DPAC,
\url{https://www.cosmos.esa.int/web/gaia/dpac/consortium}). 
Funding
for the DPAC has been provided by national institutions, in particular
the institutions participating in the {\it Gaia} Multilateral Agreement.
Support to this study has been provided by PRIN-INAF2014, "EXCALIBUR'S" (P.I. G. Clementini), 
from the Agenzia Spaziale Italiana (ASI) through grants ASI I/058/10/0 and ASI 2014-
025-R.1.2015 and by Premiale 2015, ``MITiC" (P.I. B. Garilli). G.C. thanks The Carnegie Observatories visitor programme for support as science visitor.

%%%%%%%%%%%%%%%%%%%%%%%%%%%%%%%%%%%%%%%%%%%%%%%%%%



%\bibliographystyle{mnras}


%%%%%%%%%%%%%%%%%%%%%%%%%%%%%%%%%%%%%%%%%%%%%%%%%%

%%%%%%%%%%%%%%%%% APPENDICES %%%%%%%%%%%%%%%%%%%%%

%\appendix

%%%%%%%%%%%%%%%%%%%%%%%%%%%%%%%%%%%%%%%%%%%%%%%%%%

% Don't change these lines
\bsp	% typesetting comment
\label{lastpage}
\end{document}